\documentclass[11pt]{article}
\usepackage[utf8]{inputenc}
\usepackage{amsmath}
\usepackage[normalem]{ulem}
\usepackage{amsfonts}
\usepackage{xcolor}
\usepackage{graphicx}
\usepackage[onehalfspacing]{setspace}
\usepackage[a4paper,margin=1.25in]{geometry}
\usepackage{amsthm}
\usepackage[longnamesfirst]{natbib}
\usepackage[colorlinks,linkcolor=links,citecolor=cites,urlcolor=MyDarkBlue,backref=page]{hyperref}

\usepackage{afterpage}
\usepackage{pdflscape}
\usepackage{makecell}

\usepackage[bottom]{footmisc}
\usepackage{parskip}
\usepackage{enumitem}
\usepackage[section]{placeins}

\usepackage{eurosym}

\definecolor{MyDarkBlue}{rgb}{0,0.08,0.45}
\definecolor{cites}{HTML}{1a663b}
\definecolor{links}{HTML}{1a663b}
\definecolor{MyLightMagenta}{cmyk}{0.1,0.8,0,0.1}
\usepackage{subfig}
\usepackage[justification=centering]{caption}

\begin{document}
\title{
    Managing Congestion in Two-Sided Platforms:\\The Case of Online Rentals\thanks{
        This draft: August 22, 2023.  We thank Jeremy Fox, Yu-Wei Hsieh, Xuan Teng and seminar participants at UC Irvine, the 22nd CEPR/JIE Conference on Applied Industrial Organization, the BDS-Tinbergen Institute ``Search and Markets'' Conference, the 2023 Annual Workshop on Applied Economics, the UW Optimal Transport and Econometrics Conference, and Caltech's Applied Micro Lunch for helpful comments. Zion Chang provided research assistance.}}
\date{\vspace{0.5cm}\textbf{Preliminary draft: {\em Comments welcome}}}
\author{
    Caterina\\Calsamiglia\thanks{
        ICREA, IPEG and IZA; \href{mailto:caterina.calsamiglia@barcelona-ipeg.eu}{caterina.calsamiglia@barcelona-ipeg.eu}} 
    \and 
    Laura\\Doval\thanks{
        Columbia University; \href{mailto:laura.doval@columbia.edu}{laura.doval@columbia.edu}} 
    \and 
    Alejandro\\Robinson-Cort\'es\thanks{
        University of Exeter; \href{mailto:a.robinson-cortes@exeter.ac.uk}{a.robinson-cortes@exeter.ac.uk}} 
    \and 
    Matthew\\Shum\thanks{
        California Institute of Technology; \href{mailto:mshum@caltech.edu}{mshum@caltech.edu}}
    }
\maketitle
\begin{abstract}
    Thick two-sided matching platforms, such as the room-rental market, face the challenge of showing relevant objects to users to reduce search costs. Many platforms use ranking algorithms to determine the order in which alternatives are shown to users. Ranking algorithms may depend on simple criteria, such as how long a listing has been on the platform,  or incorporate more sophisticated aspects, such as personalized inferences about users' preferences. Using rich data on a room rental platform, we show how ranking algorithms can be a source of unnecessary congestion, especially when the ranking is invariant across users. Invariant rankings induce users to view, click, and request the same rooms in the platform we study, greatly limiting the number of matches it creates. We estimate preferences and simulate counterfactuals under different ranking algorithms varying the degree of user personalization and variation across users. In our case, increased personalization raises both user match utility and congestion, which leads to a trade-off. We find that the current outcome is inefficient as it lies below the possibility frontier, and propose alternatives that improve upon it.
\end{abstract}
\clearpage
\section{Introduction}

Two-sided matching markets have experienced a significant surge in relevance in recent years, reshaping various industries and fundamentally altering the dynamics of supply and demand. The advent of digital platforms and the proliferation of online interactions have propelled the prominence of two-sided markets to new heights. Examples include ride-sharing and transportation, dating and social networking or, as our present study, accommodation and hospitality. Successful platforms require careful analysis of factors that may affect their capacity to successfully bring together and match both sides of markets.  Various design aspects that can have an impact on platforms' effectiveness need to be continuously revisited. The present paper does exactly that; using a specific application, we present a problem and a set of potential solutions applicable to a wide set of platforms.

Given the limited amount of attention users have to process the increasing number of alternatives in online marketplaces, platforms typically employ computational algorithms that steer users towards options considered good matches for them. In this paper, we use rich data on a leading room rental platform to show how currently used algorithms may induce an unnecessary amount of congestion and, in doing so, limit a platform's capacity to exhaust the set of potential matches in a market.

An unintended consequence of the steering algorithms employed by platforms is that they may create additional unnecessary congestion in the market. As in many other dynamic platforms, in the platform we study an algorithm is implemented to prioritize recent and reliable options for users to be matched with. The algorithm only generates a partial order, which ties are then broken randomly to generate a strict ranking of all apartments. Crucially, the ties are broken using the \textit{same} random number across all users. This induces users to view, click and request a relatively small set of apartments while ignoring similarly valuable ones. We find that an algorithm meant to steer renters towards apartments offered by reliable sellers, leads to congestion in the sense that many users end up clicking on the same small set of apartments, while they would have been equally or even better off exploring alternative apartments which are ranked lower by the platform's algorithm and hence do not receive enough attention and consideration from users.      

The problem in the current algorithm comes from two main sources. The first has to do with the fact that the order in which rooms are shown greatly determines what individuals view, click and request, which affects final matches. The second is driven by the fact that different users may have different preferences and hence there is a loss from having all users search in the same order. Hence, there are at least two possible solutions. The first and computationally simpler is to have the complete rank be defined by the partial order where ties are broken using a different random number for different users. The second involves personalizing the ranks for specific subgroups to ensure that the rank responds more clearly to individual preferences.

In our paper we analyze the trade-offs from personalizing more according to individual preferences to target demand to most desired rooms versus creating more randomness in the way rooms are ranked. While highly personalized algorithms steer users to highly preferred alternatives, they may cause congestion if users have similar preferences, and may reduce it if individuals have heterogeneous preferences.  Introducing randomization to ranking algorithms alleviates congestion by steering users to different rooms. The relationship between maximal utility when matched versus congestion is the crux of our analysis: can we reduce congestion without penalising users by showing them different rooms? 

We estimate preferences and simulate counterfactuals under different ranking algorithms which vary in the degree of user variation and personalization. We find a trade-off between user match utility and congestion, as increased personalization raises both of them. Moreover, the status quo algorithm used by the platform during the period we study is inefficient as its outcome lies below the possibility frontier in utility-congestion space. We design alternative algorithms that incorporate a more efficient mix of personaliation and randomization compared to the status quo outcome. They reduce congestion without decreasing the expected match utility of the rooms users are steered to, or, conversely, increase utility without reducing congestion.

Inefficiencies arising from congestion, as well as the trade-offs faced by platforms when steering consumers to their most preferred alternatives, have been documented in a handful of online platforms: \cite{fradkin17} studies search engine design and rejection rates in Airbnb; \cite{horton19}, the effects that signaling features have on matching rates in online labor markets, and \cite{ChenHsiehLin2021}, matching patterns in an online dating platform and point out, as we do, that recommendation algorithms used by tech companies may lead to congestion. 

    From a theoretical perspective, alleviating the inefficiencies caused by congestion in two-sided markets has long been seen as one of the goals of effective market design \citep{roth97,roth08}. For example, \cite{arnosti21} study the welfare losses due to inefficient equilibrium search and screening in congested markets; \cite{immorlica21} study the benefits of directing consumer search by limiting choice, which reduces congestion and the cannibalization of ``low type'' sellers' demand  by ``high types'' ones; \cite{ashlagi22} study the trade-offs between expanding choice and increasing matching when designing recommendation algorithms in two-sided platforms; \cite{mekonnen23} shows that there is more congestion when preferences are common and quality is revealed with directed search than in random matching. The main theoretical insight of our paper relates to how the effects of congestion are mediated by the differentiation present in consumer preferences, which is evidence of how optimal market design depends on market-specific features. In our case, the trade-off between utility and congestion is driven by the amount of horizontal and vertical differentiation in the market. 

    A literature parallel to the papers above studies the effects of algorithms on consumer search in online marketplaces \citep[e.g.,][]{koulayev14,HodgsonLewis2020}. While the focus of our analysis is not on modeling optimal consumer search paths, our empirical findings echo the findings in this literature, especially in relation to ranking algorithms.\footnote{As we discuss below, we model consumer clicking and requesting choice conditional on search. While this implies making the assumption that consumer search is invariant to the ranking algorithm when conducting counterfactuals, we view this favorably compared to making stronger structural assumptions in relation to consumer search behavior.} The importance of rankings in online search have been widely documented \cite[e.g.,][]{ghose14,chen17,dang22}. For example, a common finding in hotel booking platforms is that presenting the best alternatives at the top reduces the search costs associated with finding the right product \citep{delossantos17,Ursu2018}. 
 However, the platforms studied in these papers are largely many-to-one markets in which product capacity is large enough for any number of users to be matched to the same product; in these markets, congestion is not an issue, leading to design implications which differ greatly from those explored in this paper. 

    In the next subsection (\autoref{sec:motivating}), we present descriptive evidence to motivate our analysis. In \autoref{sec:model}, we present our model, estimation strategy, and a description of our empirical specification. In \autoref{sec:est_results}, we discuss the estimation results, and, in \autoref{sec:counterfactuals}, we study counterfactual exercises. We conclude in \autoref{sec:conclusion}. In \autoref{sec:appendix}, we report additional analyses and tables which we refer to throughout the text.

\subsection{Congestion: motivating evidence\label{sec:motivating}}

Our platform is one of the largest peer-to-peer rental markets in the world for medium-term rooms and apartments, active in many of the biggest cities in Europe, the Americas, and Asia. At the time period we are analyzing it differentiated itself from other competing platforms such as Airbnb by renting rooms for a longer term, at least a month, targeting exchange students and workers.
    Our platform operates via a website or mobile app in which users interested in rentals submit searches and are shown available apartments listed by landlords. We observe the sequence of clicks to view rooms and submit requests to landlords done by each user. Our platform simply brings together renters and landlords, without always participating in the booking process; in that sense our study platform has more in common with Craigslist than Airbnb, and our dataset contains limited information on what happens after a room is requested by a user. Thus the empirics in this paper focus on users' click and request behavior, for which we have complete data.
    
    Our dataset consists of 100\% of all the listed rooms and searches for Barcelona, a major city for business and university studies in Europe. In presenting search results to users, our platform ranks rooms according to an algorithm that uses several features---including the date the room was posted, and whether the room is associated with a ``registered'' landlord  who has linked their bank account on the platform---to determine the rank. A room's rank determines the position in which it is shown in the page of search results. The established criteria only determine a partial order across rooms that is subsequently randomized to generate a complete order. A key feature of this algorithm is that the {\em same} randomization is used to break ties between rooms across all renters, so that room rankings are invariant across renters.\footnote{We use user, renter, and tenant, interchangeably.} Hence, while renters may be shown different sets of rooms depending on their search criteria, the rooms will be shown {\em in the same order} to all renters at a given point in time.   
    
\begin{figure}
        \centering
        \subfloat[Share of clicks]{
        \includegraphics[width=0.45\textwidth]{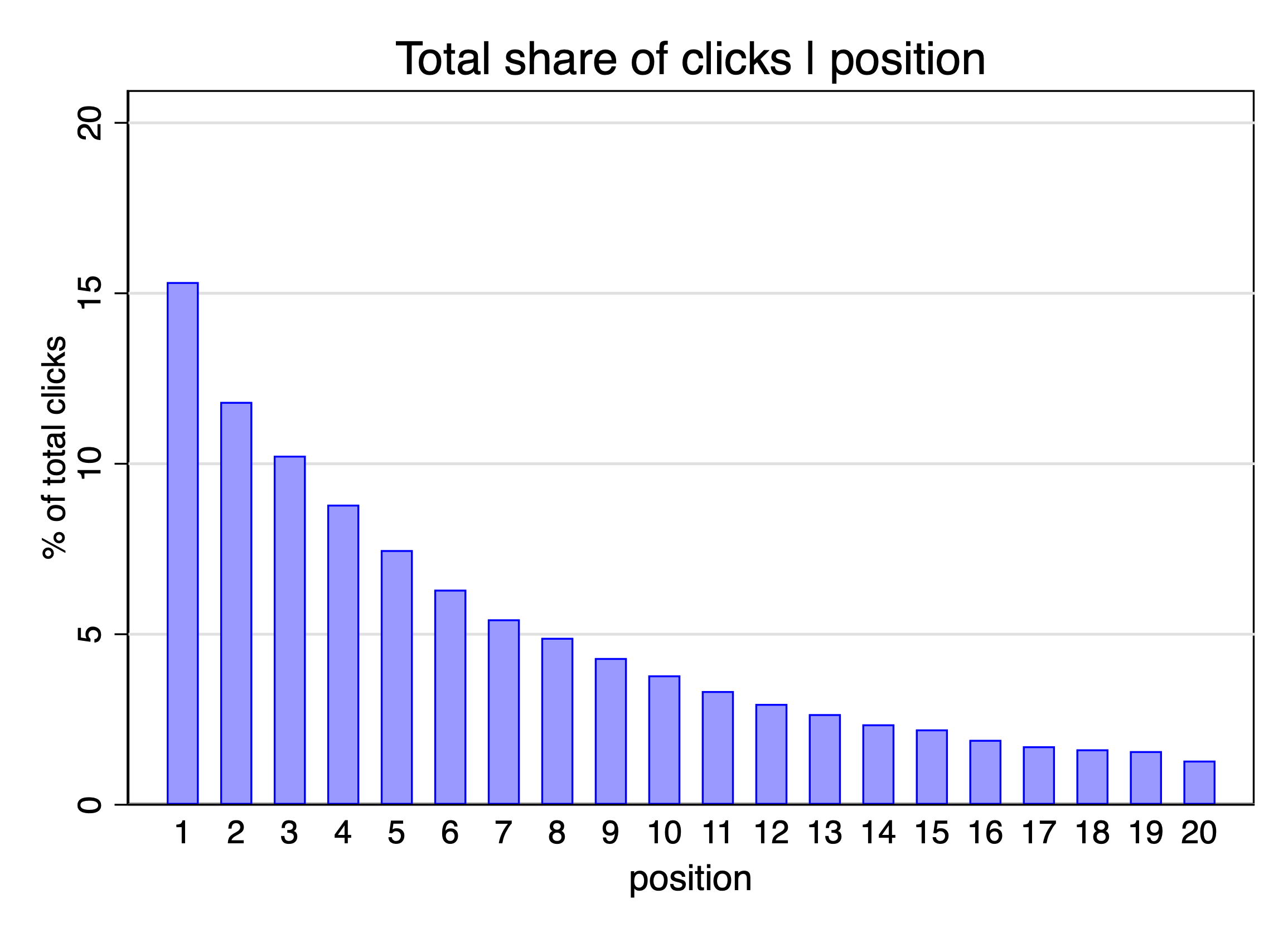}\label{fig:1a-left}}
        \hspace{1cm}
        \subfloat[Share of requests]{\includegraphics[width=0.45\textwidth]{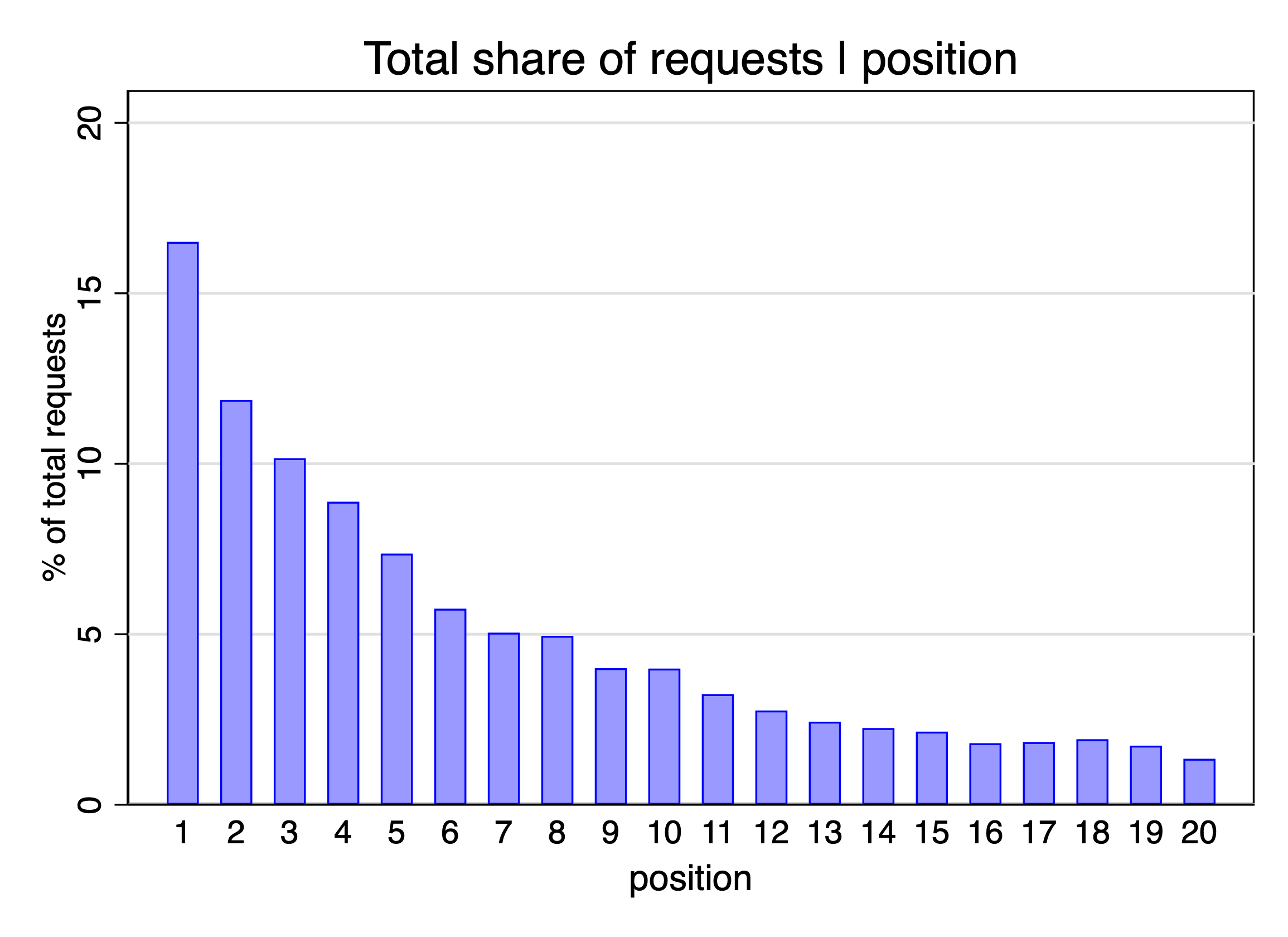}\label{fig:1b-right}}
        \caption{Concentration of clicks and requests by position\label{fig:1}}
    \end{figure}

\begin{figure}
        \centering
        \subfloat[$\mathbb{P}$(Click)]{\includegraphics[width=0.6\textwidth]{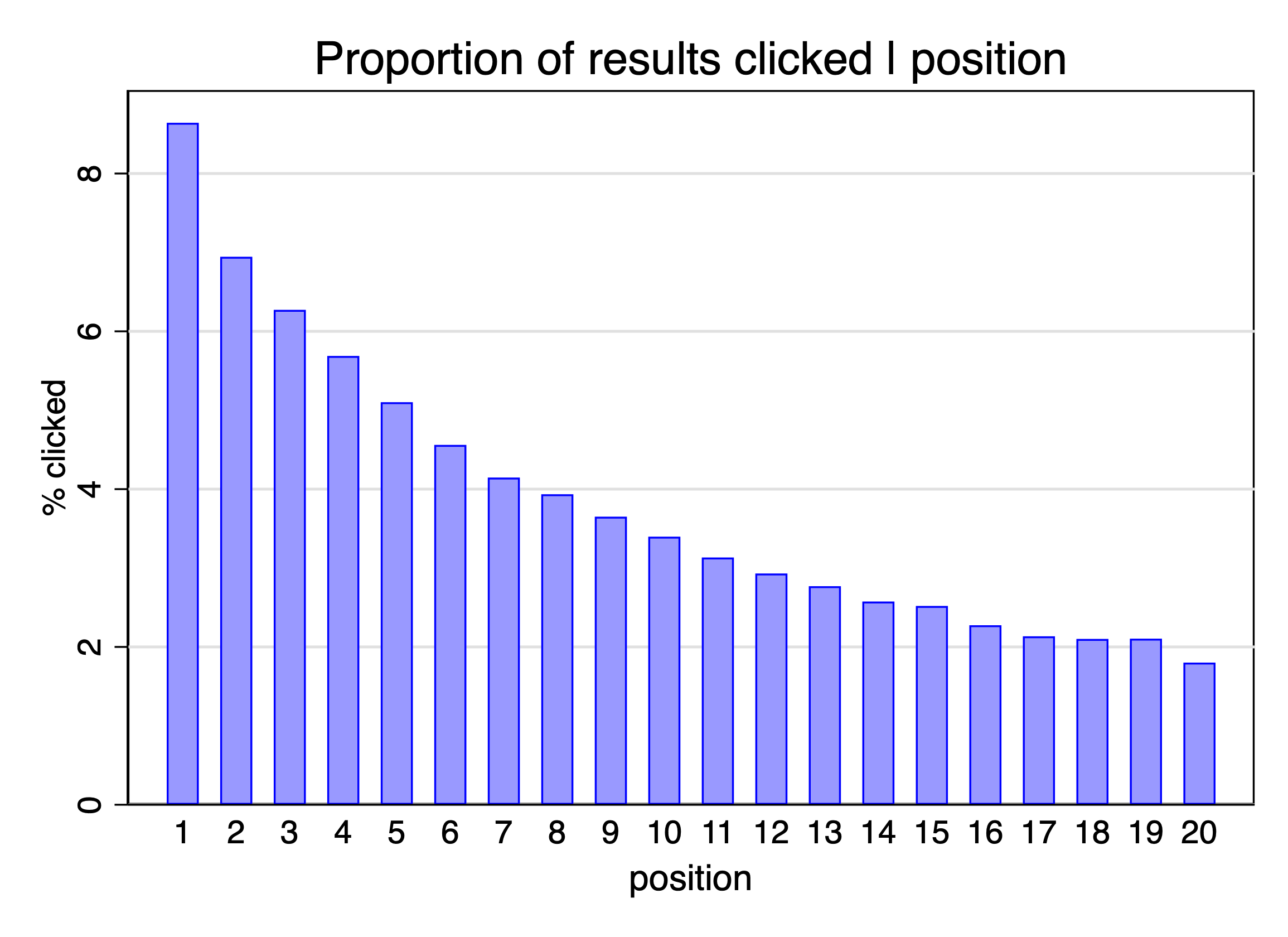}\label{fig:2a-top}}
        \\
        \subfloat[$\mathbb{P}$(Request)]{\includegraphics[width=0.6\textwidth]{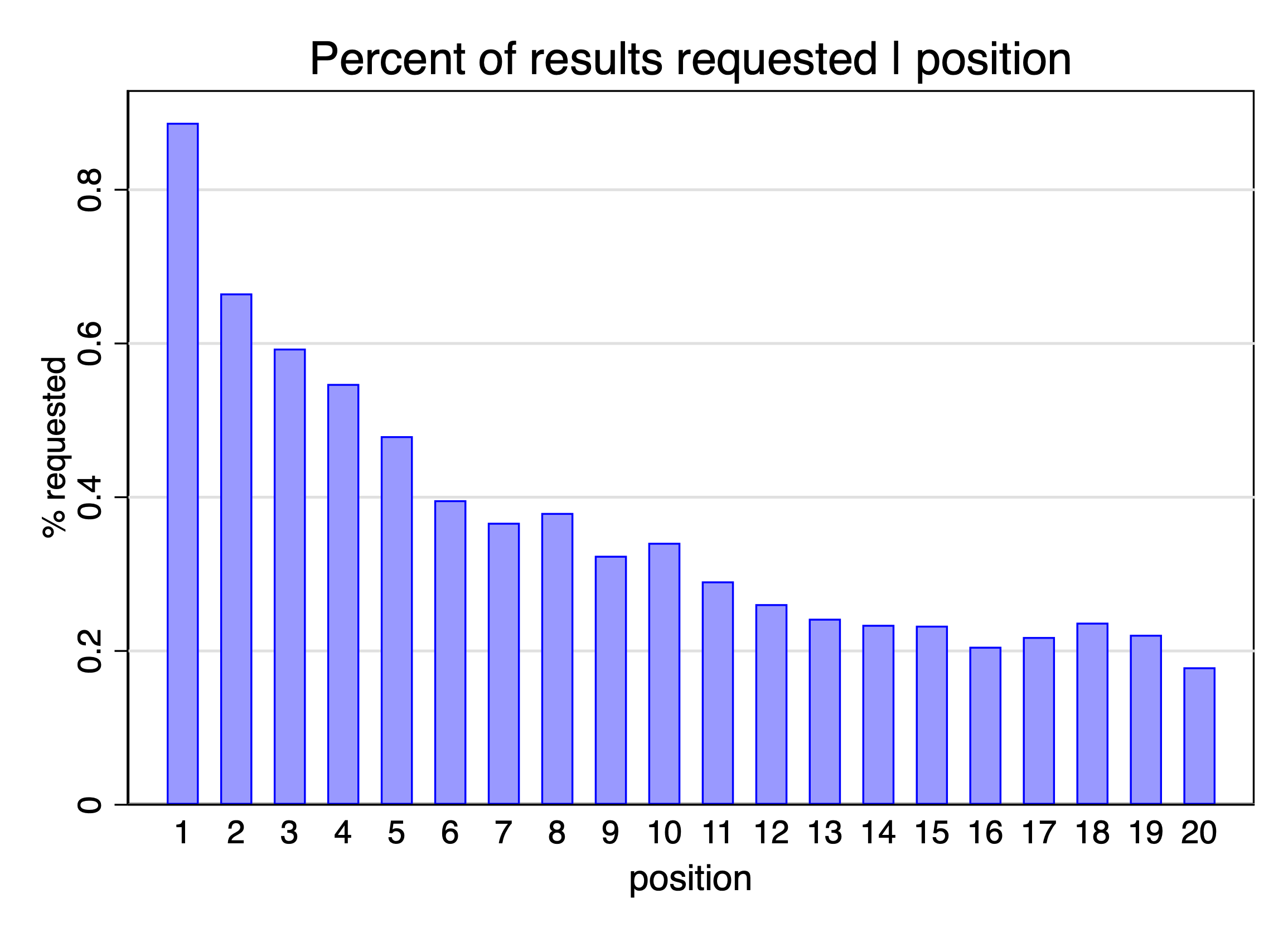}\label{fig:2b-center}}
        \\
        \subfloat[$\mathbb{P}$(Request $\!\mid\!$ Click=1)]{\includegraphics[width=0.6\textwidth]{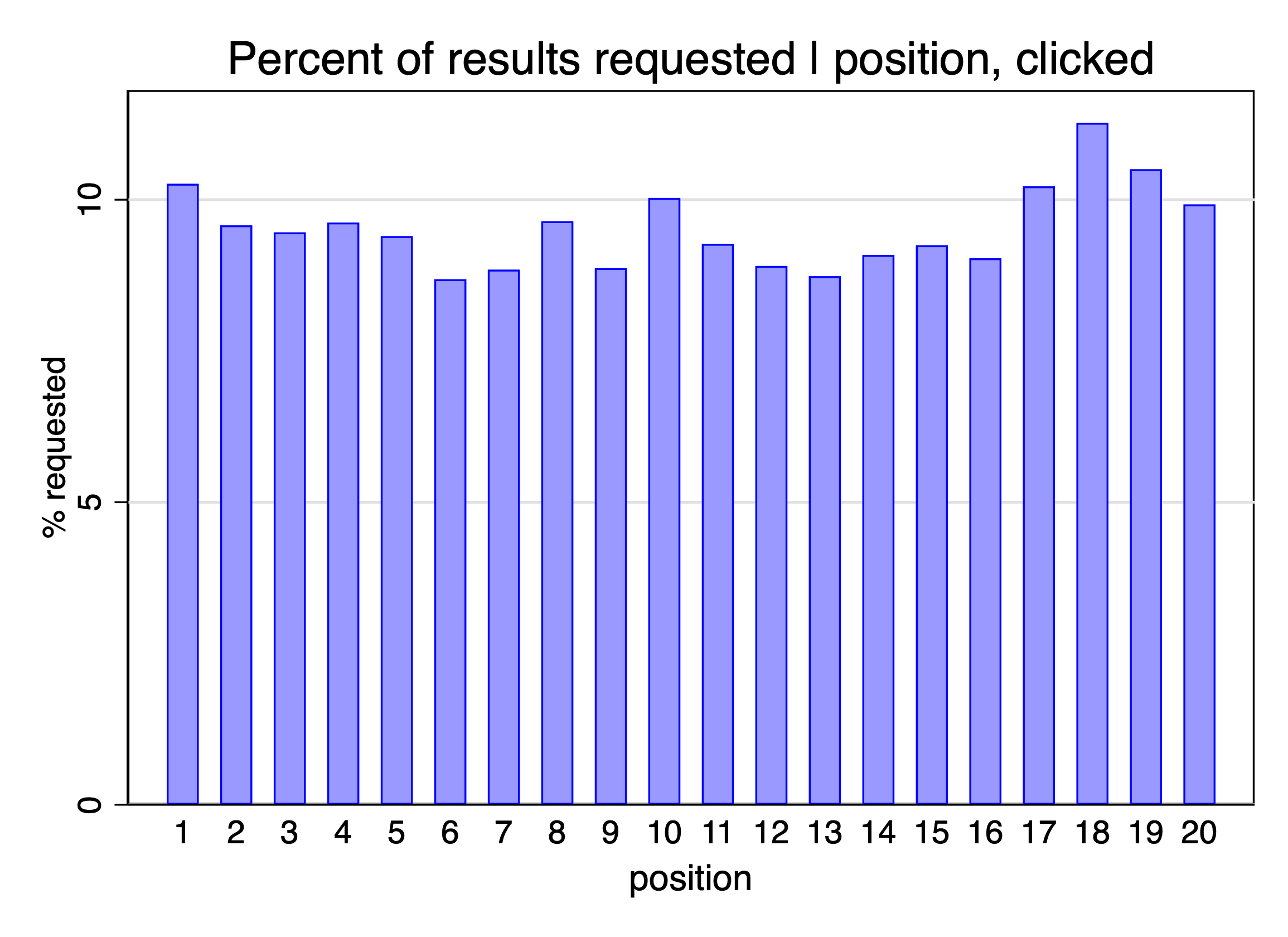}\label{fig:2c-bottom}}
        \caption{Probability of click and request by position\label{fig:2}}
    \end{figure}

        \autoref{fig:1} shows that clicks (\autoref{fig:1a-left}) and room requests (\autoref{fig:1b-right}) are concentrated on rooms shown to users at top positions. Specifically, we see that over 15\% of all clicks and requests in the data are made to rooms shown to users at the first position. This percentage falls to around 4\% for the tenth position, and asymptotes to around 2\% for the 20-th position. This evidence suggests that rankings matter to renters' behavior. Indeed, the top two graphs of \autoref{fig:2} show that  rooms shown in higher positions are more likely to be clicked on (\autoref{fig:2a-top}) and requested (\autoref{fig:2b-center}).

    However, the bottom graph (\autoref{fig:2c-bottom}) offers an interesting counterpoint. Conditional on being clicked, rooms in higher positions are not more likely to be requested than those in lower positions.\footnote{This is reminiscent of \cite{Ursu2018}, who finds a similar pattern analyzing data of Expedia.} Specifically, conditional on having been clicked on, a top-ranked room and the 20-th ranked room are both requested with a probability close to 0.10. This suggests that the rooms' page ranking is to some extent orthogonal to the users' preferences over rooms, as it affects only whether users click on a room (which arguably depends on users' attention), but not whether they request a room (which depends on their preferences over the room's characteristics). Since rooms have to be clicked on before they can be requested, this also suggests that the concentration at the top positions observed in \autoref{fig:1} might be driven more by clicking behavior and attention than preferences.

    \begin{figure}
        \centering
        \subfloat[Searches\label{fig:3a-top}]{\includegraphics[width=0.6\textwidth]{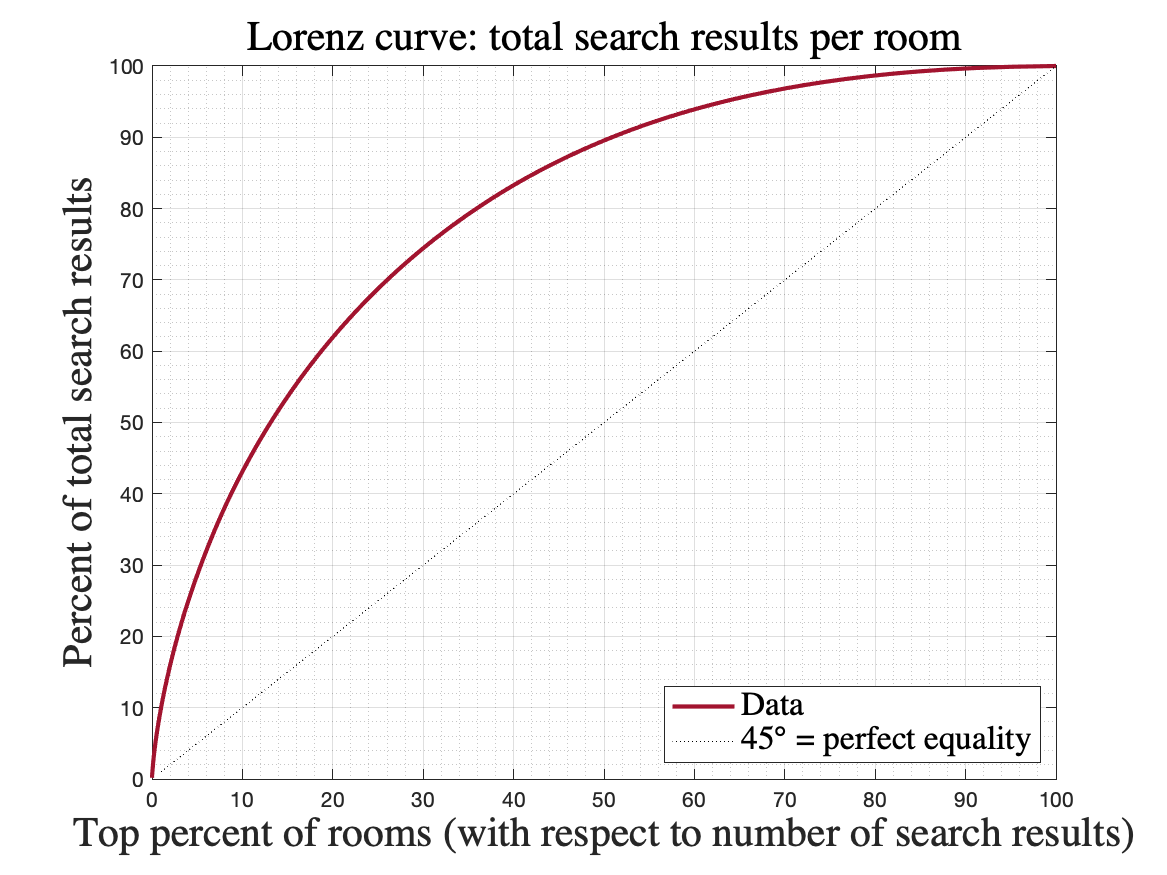}}
        \\
        
        \subfloat[Clicks\label{fig:3b-center}]{\includegraphics[width=0.6\textwidth]{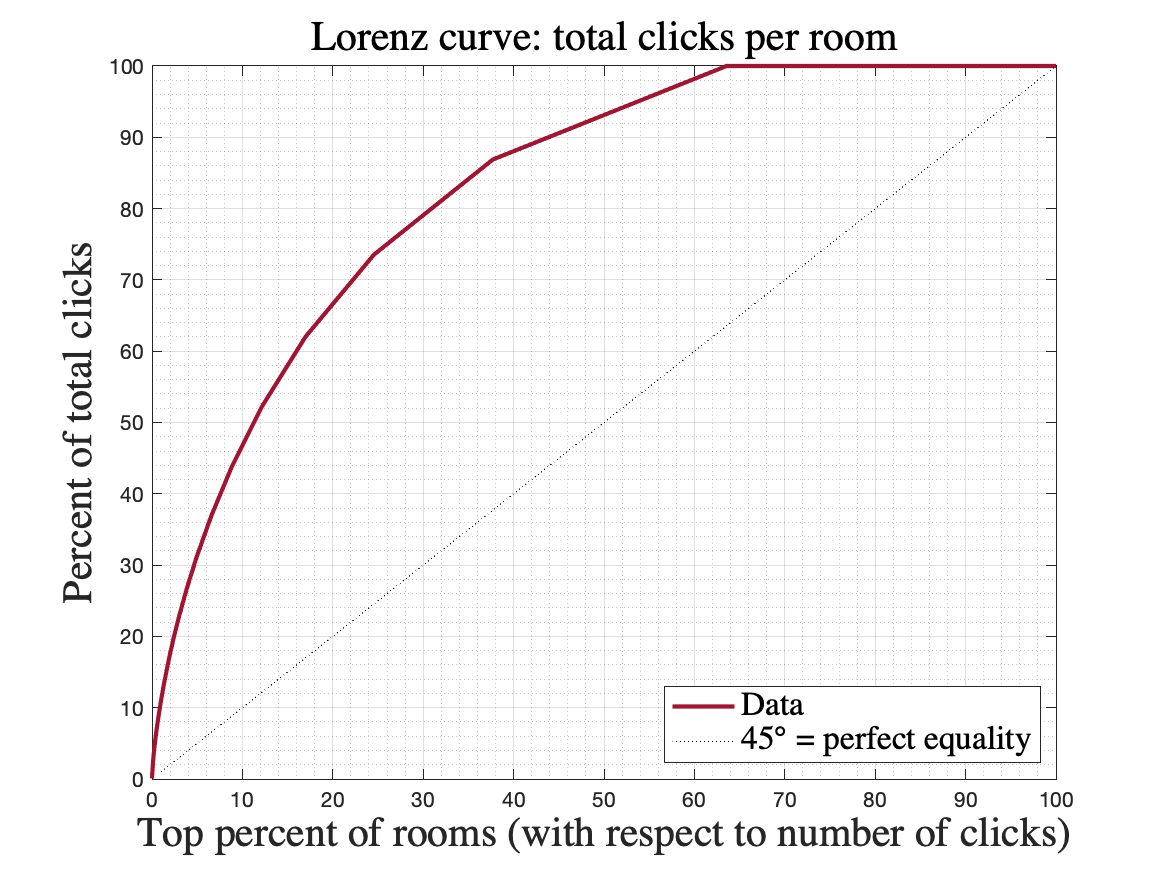}}
        \\
        
        \subfloat[Requests\label{fig:3c-bottom}]{\includegraphics[width=0.6\textwidth]
        {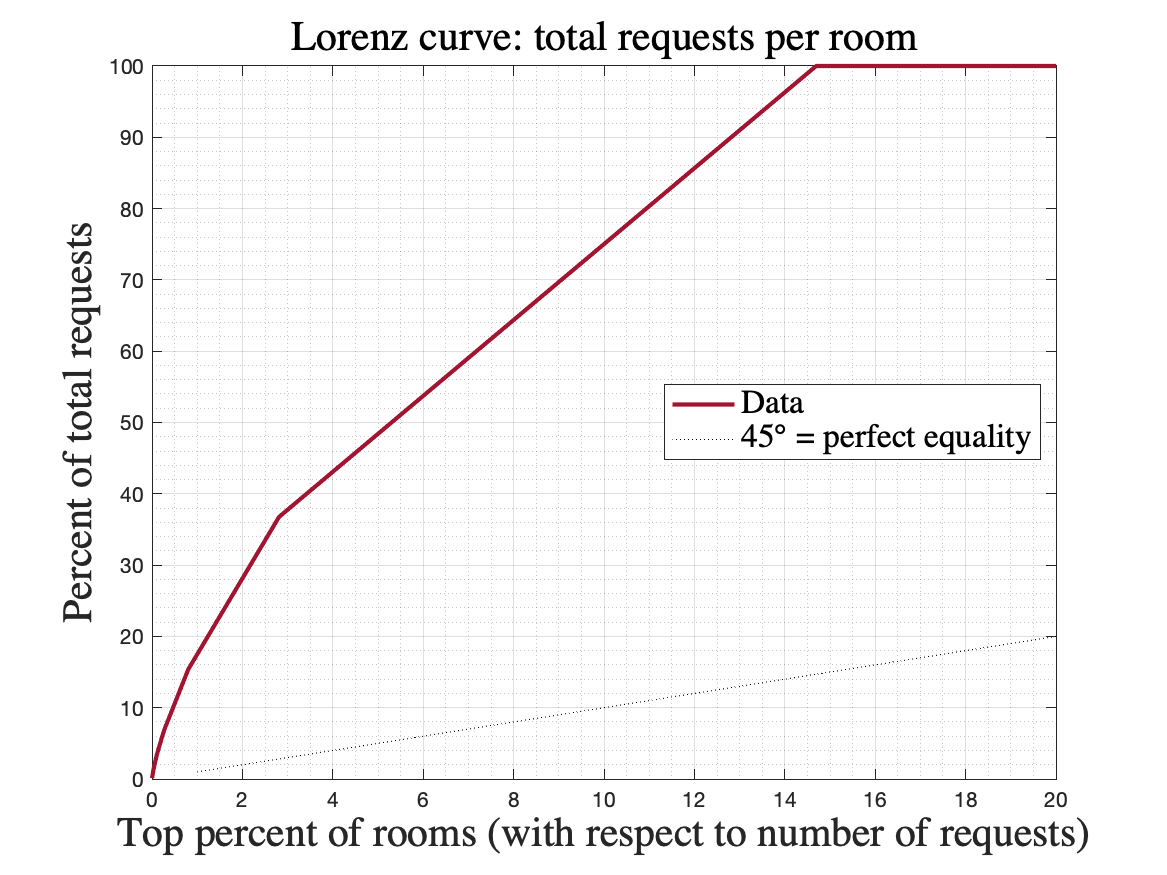}}
        
        \caption{Concentration of searches, clicks and requests\label{fig:3}}
    \end{figure}  

    \autoref{fig:3} presents initial evidence relating to the congestion on the platform across rooms. To illustrate congestion, we use Lorenz curves, which measure the degree of concentration, or congestion, across rooms in terms of the frequency with which they appear as search results, and are clicked on or requested. Higher congestion bends the Lorenz curve towards the plot's upper-left corner and indicates that the frequency is nonuniform across rooms; lower congestion moves the Lorenz curve towards the 45-degree line, indicating that the distribution across rooms is closer to uniform.
        
        The top graph in \autoref{fig:3} shows the Lorenz curve for search results, which illustrates a high degree of congestion: the 20\% of rooms that appear most often in users' search results (on the $x$-axis) accumulate around 65\% of the total number of search results in the data (on the $y$-axis). This congestion only worsens when we consider clicked rooms, as from the middle graph (\autoref{fig:3b-center}) we see that the top 20\% of rooms in terms of clicks account for almost 70\% of the total clicks.  Finally, the bottom graph (\autoref{fig:3c-bottom}) shows that the top 20\% of rooms in terms of requests make up the entirety (100\%) of all room requests.\footnote{The dataset used in the analysis contains a random sample of 10\% of the users in the market for rooms in Barcelona. We specify in detail how we construct the sample in \autoref{sec:empirical_spec}.}   
    
    This suggests substantial congestion as the universe of clicks and requests are highly concentrated on around 20\% of the rooms. At the same time, 80\% of the rooms never receive a single request, which raises the possibility of allocative gains from reducing congestion. Since a room can be rented to at most one user, all else equal, renters are better off requesting rooms that receive fewer requests from other renters. These are topics which will guide our subsequent analysis.

\begin{figure}
        \centering
        \includegraphics[width=0.7\textwidth]{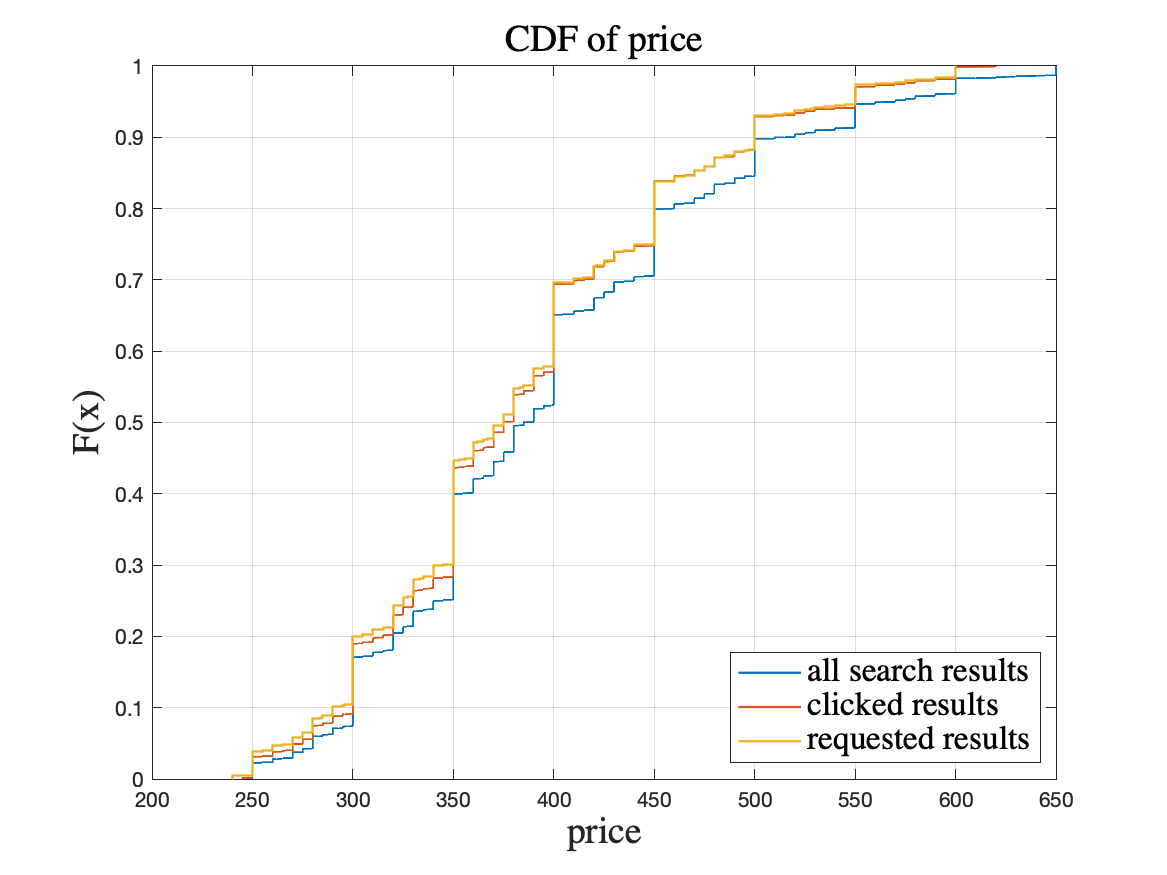}
        \caption{Price sensitivity during Customer Search Journey\label{fig:4}}
    \end{figure}

    Finally, \autoref{fig:4} illustrates users' price sensitivity. We plot CDFs of room prices across all search results (blue), search results that are clicked on (red), and search results that are requested (orange). We see that the CDF of prices across all search results stochastically dominates the other two; users tend to click and request cheaper rooms. The figure also shows, albeit in a lower magnitude, that the distribution of prices across clicked rooms stochastically dominates that of rooms that are requested. In other words, as the consumer search journey progresses, the rooms users browse, click on, and eventually request, become cheaper. As we would expect, this shows that users are price sensitive, and may even update their evaluations of rooms after clicking on them and seeing their listings in more detail.

\section{Model}\label{sec:model}

\autoref{sec:model} introduces the setup and notation which underlies our analysis. 
    Users (or renters) are indexed by $i\in N=\{1,\dots,n\}$. Each user is associated to a vector of observable characteristics $Z_i$. User $i$ conducts searches $S\in \mathcal{S}_i$, where $|\mathcal{S}_i|=n_i$ is the number of searches conducted by user $i$. A search $S$ is a collection of search results $s\in S=\{1,\dots,n_{S}\}$, where $n_S$ is the number of search results in search $S$.
    
    Each search result $s$ is associated to a vector of observable characteristics, $X_s=(X_{1s},X_{2s},pos_s)$, where $X_{1s}$ are the characteristics users observe in the page of search results, such as the price and the location of a room, $X_{2s}$ the ones they observe only upon clicking on the search result, such as the amenities of a room, and $pos_s$ is the position the room is shown on the page of search results.\footnote{Specifically, a search $S$ is a collection of pages, each containing several search results. However, for simplicity we abstract from the page a search result is shown in, and just take into account the position it is shown in within the corresponding page. This should not have an impact on our counterfactual analysis if we assume that the changes in the ranking algorithm do not change the number of pages explored by users.} The decision variables observed in the data are:
    \begin{align}
        k_{is}&=1\{\text{user $i$ clicks on search result $s$}\}\\
        r_{is}&=1\{\text{user $i$ requests search result $s$}\}
    \end{align}
    We omit the dependence of $k_{is}$ and $r_{is}$ on $S\in\mathcal{S}_i$ for simplicity. Since to request a room one needs to click on it first, we have $r_{is}=1\Rightarrow k_{is}=1$. Finally, we index rooms by $j\in J$, where $j_s$ denotes the room associated to search result $s$.

\subsection{Users' clicks and room requests: A multinomial choice framework}

Next we describe the framework we use to model users' observed click and room requests. 
    The utility of requesting search result $s$ in a generic search $S$ made by user $i\in N$ is given by:
    \begin{equation}\label{eq:request-utility}
        U_{is} = X_{1s}\beta_1+X_{2s}\beta_2+(X_{2s}\otimes Z_i)\beta_{XZ}+\varepsilon^r_{is},
    \end{equation}
    where $\varepsilon^r_S=(\varepsilon^r_{is})_{s\in S}$ is an iid vector of extreme-value type 1 random utility shocks and $X_{2s}\otimes Z_i$ is a shorthand for interactions between $X_{2s}$ and $Z_i$ (which may not be exhaustive). For example, $X_{2s}\otimes Z_i$ includes whether user $i$ has the gender, age, or occupation preferred by the landlord of the listing in search result $s$.
    
    The request data reveals a partial ordering of the utility of rooms conditional on being clicked, where rooms that are requested have a utility higher than those that are clicked but not requested. For example, suppose that a user clicks on four rooms (search results) in a search, 1, 3, 4, and 7, with utilities $U_1$, $U_3$, $U_4$, and $U_7$, respectively. If they request, say, rooms 1 and 4, the request data reveals $\min\{U_1,U_4\}>\max\{U_3,U_7\}$. 
    Essentially, in the estimation we allow the users to choose more than one item among their choice sets, and impose only that users obtain higher utility from the chosen options than the non-chosen options, while being agnostic about users' ranking among the chosen (and non-chosen) options. Given the extreme-value specification of the utility shocks, our model takes the form of a rank-ordered logit model with ties \citep{beggs81,allison1994}.\footnote{We have also estimated semiparameteric versions of this model \citep{Fox2007,yan17}, and our findings are qualitatively similar.}
        
    The propensity to click on a room depends on the position the room is shown in  as a search result and on its expected utility conditional on the information the user has before clicking. Formally, the propensity to click on a search result $s$ in a generic search $S$ of user $i\in N$ is determined by:
    \begin{equation}\label{eq:click-utility}
        I_{is} = g(pos_{is})\beta_{pos}+\mathbb{E}\left[U_{is}\mid X_{1s},pos_{is}\right]\beta_U+pos_{is}\times\mathbb{E}\left[U_{is}\mid X_{1s},pos_{is}\right]\beta_{pos\times U}+\varepsilon^k_{is},
    \end{equation}
    where $\varepsilon^k_S=(\varepsilon^k_{is})_{s\in S}$ is an iid vector of extreme-value type 1 random shocks, $g(\cdot)$ is a known function (which we specify as a squared polynomial with point-masses at the first, second, and third positions), and $\mathbb{E}\left[U_{is}\mid X_{1s},pos_{is}\right]$ is the room's expected utility conditional on the available information. Namely,
    \begin{align}
        \mathbb{E}\left[U_{is}\mid X_{1s},pos_{is}\right] &=X_{1s}\beta_1+\mathbb{E}\left[X_{2s}\mid X_{1s},pos_{is}\right]\beta_{2}+\mathbb{E}\left[X_{2s}\otimes Z_i\mid X_{1s},pos_{is}\right]\beta_{XZ}.\label{eq:cond_utility}
    \end{align}
  As with requests, our model for clicks has the structure of a rank-ordered logit with ties. 

\subsection{Estimation}

    There are two stages in users' choice behavior. First, a user {\em clicks} a room from her search results. Second, after learning more about its features, she may decide to {\em request} the room by contacting the landlord. Analogously, there are two sequential stages in our estimation procedure, corresponding to users' click and request decisions. Proceeding backwards, we start by describing the estimation of the model for the request decision, and then describe how this feeds into our estimation of the earlier click decision.
    
    In the first step, we estimate the parameters in the utility equation \eqref{eq:request-utility} via maximum likelihood: $\hat{\beta}_1$, $\hat{\beta}_2$, and $\hat{\beta}_{XZ}$. After estimating these parameters, we project the covariates in $X_{2s}$ onto $X_{1s}$ and $pos_{is}$ to estimate the expected requested utility. The estimated expected request utility is given by:
    \begin{equation}\label{eq:estimated_util}
        \hat{U}_{i}(X_s) = X_{1s}\hat{\beta}_1+\hat{X}_{2s}\hat{\beta}_2+(\hat{X}_{2s}\otimes Z_i)\hat{\beta}_{XZ},
    \end{equation}
    where $\hat{X}_{2s}$ are the fitted values of linear projections (in which we include numerous interactions to provide enough flexibility). Equation \eqref{eq:estimated_util} is the estimated version of \eqref{eq:cond_utility}.
        
    Next we proceed to estimate the click decision model, where we use the estimated expected request utility $\hat{U}_{i}(X_s)$ as a covariate and interact it with $pos_{is}$ (see equation \eqref{eq:click-utility}). The parameters in the click equation \eqref{eq:click-utility} are readily estimated via maximum likelihood: $\hat{\beta}_{pos}$, $\hat{\beta}_{U}$, and $\hat{\beta}_{pos\times U}$.

\subsection{Empirical specification and summary statistics}\label{sec:empirical_spec}

In the empirical specification, we include the following variables:
    \begin{itemize}
        \item $X_{1s}=$ price, number of tenants in room (may be missing in which case we include a dummy), and location (we split Barcelona into six districts);
        \item $X_{2s}=$ room amenities (AC, Balcony, Dishwasher, Doorman, Elevator, Exterior View, Heating, Smoker Friendly, TV, Terrace), number of days the room has been published in the platform, and dummies indicating whether the user has the age, gender, and occupation (student or worker) preferred by the landlord.
    \end{itemize}
        
    Our dataset consists of a random sample comprising 10\% of all the users who used the platform in a two-year period (January 2018 to February 2020). For each of these users, we observe the complete sequence of searches, clicks, and requests. There exists vast variation across users and searches in the data, with many users searching all over the globe and others searching without submitting any requests. In order to include in our analysis only the relevant choices made by users who are likely to be considering renting an apartment in Barcelona, we restrict our sample as follows. First, we say that a search is in Barcelona if at least 75\% of its search results are in Barcelona, where a search result is said to be in Barcelona if the room's location is at most 30km away from Barcelona's city center. We include in the estimation sample the searches in Barcelona made by users who (i) predominantly search in Barcelona (at least 75\% of their searches are in Barcelona), and (ii) made at least one request to a room in Barcelona during the sample period.
    
    \autoref{tab:1} provides the size of our estimation sample and descriptive statistics of the users, rooms, searches, clicks, and requests in it. The sample contains 1,202 users. On average, users are 29 years old; 51\% of them are females, 35\% are students, and 81\% are employed. There are 45,462 rooms, of which 80\% specify they prefer renters within a specific age bracket (on average, they prefer renters that are between 20 and 36 years old). 27\% of rooms also specify a preferred gender for renters, of which 23\% prefer female renters and 4\% male renters. Finally, 25\% of rooms specify a preferred occupation for renters, of which 22\% prefer \textit{not} renting to students, and 3\% prefer to rent to students \textit{only}.
    
    In terms of congestion, while all rooms appear at least once as a search result in the sample, only 64\% of them receive at least one click, and 15\% receive at least one request (see \autoref{fig:3} for a full picture of the concentration of searches, clicks, and requests in the sample). On average, each user conducts 144 searches and browses over 1,700 search results. Searches contain an average of 12 search results, with a maximum of 20. This adds up to a total of over 2 million search results. On average, each user clicks on 75 rooms and submits requests to 7 rooms.

    \begin{table}[h!]
        \centering\small
        \caption{Sample description\label{table_1}}
        \begin{tabular}{ll}\hline\hline
         &  \\
        Number of users (renters) & 1,202 \\
        \hspace{0.5cm} Average age & 29.44 \\
        \hspace{0.5cm} Are females (\%) & 51.4 \\
        \hspace{0.5cm} Are students (\%) & 34.61 \\
        \hspace{0.5cm} Are employed (\%) & 81.03 \\
         &  \\\hline
         &  \\
        Number of rooms & 45,462 \\
        \hspace{0.5cm} Specify minimum tenant age (\%) & 80.27 \\
        \hspace{0.9cm} Average minimum tenant age & 20.57 \\
        \hspace{0.5cm} Specify maximum tenant age (\%) & 80.27 \\
        \hspace{0.9cm} Average maximum tenant age & 36.06 \\
        \hspace{0.5cm} Specify a preferred tenant gender (\%) & 27.2 \\
        \hspace{0.9cm} Prefer female tenants (\%) & 23.13 \\
        \hspace{0.9cm} Prefer male tenants (\%) & 4.07 \\
        \hspace{0.5cm} Specify a preferred tenant occupation (\%) & 25.4 \\
        \hspace{0.9cm} Prefer no student tenants (\%) & 22.7 \\
        \hspace{0.9cm} Prefer not employed tenants (\%) & 2.68 \\
        \hspace{0.5cm} Appear in at least one search result (\%) & 100 \\
        \hspace{0.5cm} Are clicked on at least one (\%) & 63.56 \\
        \hspace{0.5cm} Are requested at least one (\%) & 14.69 \\
         &  \\\hline
         &  \\
        Total number of searches & 173,223 \\
        \hspace{0.5cm} Average per user & 144.11 \\
         &  \\\hline
         &  \\
        Total number of search results & 2,066,147 \\
        \hspace{0.5cm} Average per user & 1,718.92 \\
        \hspace{0.5cm} Average per room & 45.45 \\
        \hspace{0.5cm} Average per search & 11.93 \\
         &  \\\hline
         &  \\
        Total number of clicks & 89,624 \\
        \hspace{0.5cm} Average per user & 74.56 \\
        \hspace{0.5cm} Average per room & 1.97 \\
        \hspace{0.5cm} Average per room $\mid$ $\geq 1$ & 3.1 \\
        \hspace{0.5cm} Average per search & 0.52 \\
        \hspace{0.5cm} Prob.\ search result is clicked & 4.3\% \\
         &  \\\hline
         &  \\
        Total number of requests & 8,542 \\
        \hspace{0.5cm} Average per user & 7.11 \\
        \hspace{0.5cm} Average per room & 0.19 \\
        \hspace{0.5cm} Average per room $\mid$ $\geq 1$ & 1.28 \\
        \hspace{0.5cm} Average per search & 0.049 \\
        \hspace{0.5cm} Prob.\ click is requested & 9.5\% \\
        \hspace{0.5cm} Prob.\ search result is requested & 0.41\% \\
         &  \\\hline\hline
        \multicolumn{2}{p{9cm}}{\scriptsize\textit{Notes:} Table reports sample sizes and descriptive statistics of the users (renters), rooms, searches, search results, clicks, and requests in the estimation sample.\label{tab:1}}
        \end{tabular}
    \end{table}
    
    \autoref{tab:2} reports summary statistics of room characteristics. The left panel (first five columns) reports summary statistics across all search results in the sample.\footnote{We provide characteristics at the level of search results since this is the unit of observation in the analysis.} The right panel (last three columns) reports mean differences across: (a) search results that are clicked on and all search results (clicked -- all); (b) search results that are requested and all search results (requested -- all), and (c) search results that are requested and clicked on (requested -- clicked). Analyzing these differences in means offers a quick insight into the presence of variation across the rooms that renters browse, click on, and ultimately request. Our model is designed to exploit this variation to estimate users' preferences for rooms.
    
    On average, rooms displayed as search results are around \euro 400 per month. Compared to all displayed rooms, on average, the ones that are clicked on are \euro 12 cheaper, and the ones that are requested are \euro 15 cheaper. We also see that renters tend to click on rooms that have been published for more days in the platform, but do not request them. Renters tend to browse rooms for which their personal characteristics match the landlord's preferences: 81\% of search results have the preferred gender of the user (or none), 65\% their preferred age (or none), and 90\% their preferred occupation (or none). 
    Unsurprisingly, renters are more likely to click on and request listings in which their gender, age, and occupation match the landlord's preferences. 
    There is rich variation in the data across room amenities: doorman, dishwasher, terrace, etc. In general, renters tend to click on and request rooms that have more amenities. Interestingly, users seem less inclined to click on and request rooms with AC. By contrast, having a balcony or an exterior view are amenities that renters seem to value more. In terms of location, the sample is relatively balanced across the six regions of the city we consider.\footnote{The city of Barcelona proper has 10 districts, which we group into five by merging together Sarri\`a-Sant Gervasi and Gr\`acia, Ciutat Vella and Saint Mart\'i, Sants-Montjuic and Les Corts, and  Horta-Guinard\'o, Nou Barris and Sant Andreu (into North Barcelona). Additionally, we pool together rooms that are within a 30km radius from the city center but outside the 10 districts, into ``Greater Barcelona.''} Perhaps unsurprisingly, renters seem to be more likely to click and request rooms near the city center, in Ciutat Vella, Saint Mart\'i, or L’Eixample.

    \afterpage{
    \begin{landscape}
    \begin{table}[htbp!]
    \centering\small
    \caption{Summary statistics and mean differences}
    \begin{tabular}{lcccc m{1cm}|@{\hskip 0.55cm}lll}
    \hline\hline
        & mean & sd & median & min & \multicolumn{1}{l}{max} & 
    \scriptsize\makecell[b]{ clicked \\ $-$ all }   &
    \scriptsize\makecell[b]{ requested \\ $-$ all }  &
    \scriptsize\makecell[b]{ requested \\ $-$ clicked } \\
    \hline
    &  &  &  &   &  &   &   & \\
    Price               &     400.591&     102.643&         390&           0&         750&      -12.377***&      -15.288***&       -3.740***\\
    Missing number of tenants&       0.132&       0.339&           0&           0&           1&       -0.009***&       -0.005   &      0.004   \\
    Number of tenants   &       1.898&       1.406&           2&           0&          20&      0.018***&      0.047***&      0.032** \\
    Days since first published&     100.669&     166.695&          14&           0&         744&      5.648***&       -5.017***&      -11.495***\\
    &  &  &  &   &  &   &   & \\
    User has gender preferred by landlord&       0.812&       0.391&           1&           0&           1&      0.008***&      0.034***&      0.029***\\
    User has age preferred by landlord&       0.654&       0.476&           1&           0&           1&      0.006***&      0.026***&      0.022***\\
    User has occupation preferred by landlord&       0.902&       0.297&           1&           0&           1&      0.006***&      0.013***&      0.008** \\
    &  &  &  &   &  &   &   & \\
    Has doorman         &       0.142&       0.349&           0&           0&           1&      0.003***&       -0.004   &       -0.008*  \\
    Has dishwasher      &       0.187&       0.390&           0&           0&           1&      0.003** &       -0.001   &       -0.004   \\
    Has terrace         &       0.214&       0.410&           0&           0&           1&      0.006***&      0.001   &       -0.005   \\
    Has AC              &       0.241&       0.428&           0&           0&           1&       -0.004***&       -0.017***&       -0.014***\\
    Is smoker friendly  &       0.278&       0.448&           0&           0&           1&      0.002   &      0.011** &      0.010** \\
    Has elevator        &       0.653&       0.476&           1&           0&           1&       -0.001   &      0.004   &      0.006   \\
    Has exterior view   &       0.492&       0.500&           0&           0&           1&      0.014***&      0.022***&      0.009   \\
    Has TV              &       0.614&       0.487&           1&           0&           1&      0.003*  &       -0.002   &       -0.005   \\
    Has balcony         &       0.417&       0.493&           0&           0&           1&      0.015***&      0.040***&      0.027***\\
    Has heating         &       0.470&       0.499&           0&           0&           1&      0.004** &      0.011** &      0.008   \\
    &  &  &  &   &  &   &   & \\
    In Sarri\`a Saint-Gervasi or Gr\`acia&       0.130&       0.336&           0&           0&           1&      0.007***&      0.005   &       -0.002   \\
    In Ciutat Vella or Saint Mart\'i&       0.198&       0.399&           0&           0&           1&      0.003** &      0.043***&      0.044***\\
    In L’Eixample         &       0.263&       0.440&           0&           0&           1&      0.016***&      0.036***&      0.022***\\
    In North of Barcelona&       0.124&       0.329&           0&           0&           1&       -0.022***&       -0.043***&       -0.025***\\
    In Sants-Montjuic or Les Corts&       0.160&       0.366&           0&           0&           1&       -0.008***&       -0.024***&       -0.018***\\
    In Greater Barcelona&       0.125&       0.330&           0&           0&           1&      0.003** &       -0.017***&       -0.022***\\
    &  &  &  &   &  &   &   & \\
    \hline\hline
    \multicolumn{9}{p{20cm}}{\scriptsize\textit{Notes:} the table reports summary statistics of the variables used in the model. The first five columns report the mean, standard deviation, median, minimum, and maximum across all search results in the sample. The last three columns report mean differences between (a) clicked and all search results; (b) requested and all search results; (c) requested and clicked search results. * $p<0.10$, ** $p<0.05$, *** $p<0.01$. Variables \textit{Price} and \textit{Days since first published} are winsorized on the right at the 99th percentile. See \autoref{tab:1} for sample sizes and additional descriptive statistics.\label{tab:2}}
    \end{tabular}
    \end{table}
    \end{landscape}
    }
    \clearpage

\section{Estimation Results}\label{sec:est_results}

In \autoref{sec:est_results}, we report and discuss the results of our estimation exercise. 
    \autoref{tab:req_bs} and \autoref{tab:click_bs} report the estimated parameters of the request utility and the click propensity, respectively. 
    
    First, we discuss the parameter estimates of the request utility. As \autoref{tab:req_bs} shows, we estimate a negative and statistically significant price sensitivity across multiple specifications. The coefficients are only identified up to scale in our model. To interpret their magnitudes it is convenient to rescale them by dividing our estimates by the absolute value of the price coefficient which allows us to interpret their magnitudes as willingness-to-pay in euros. For example, once normalized, the coefficient in the last specification for days since first published equals $-0.26$, which means that being listed for an additional day in the platform has an effect on the probability that a user submits a request to a room after having clicked on it equivalent to the room being \euro 0.26 more expensive. While statistically significant different from zero, the magnitude of this coefficient does not point to a large economic significance. \autoref{tab:norm_req_bs} in \autoref{appendix:more_tables} reports the normalized values of all coefficients. 
    Seen through this light, we observe that the magnitudes of the coefficients seem reasonable and economically significant, ranging from values close to zero up to a few hundred euros. For example, having the gender preferred by the landlord is valued positively at \euro 521, having a doorman is valued negatively at \euro 121, and having a balcony and heating are valued positively at \euro 113 and \euro 102, respectively. 
    In terms of statistical significance, few room characteristics appear to have a significant impact on consumers' preferences for rooms. Given the large sample size, this may be driven by limited variation in the rooms that users click on. However, as we shall see below, for counterfactuals, we focus on the overall predicted utility of rooms, rather than on specific coefficients.

    \begin{table}[h!]
    \centering\small
    \caption{Parameter estimates of request utility\label{tab:req_bs}}
    \begin{tabular}{lcccc}\hline\hline
                                             &         (1)   &         (2)   &         (3)   &         (4)   \\\hline
    Price                                    &     -0.0007** &     -0.0007** &     -0.0008***&     -0.0009***\\
                                             &    (0.0003)   &    (0.0003)   &    (0.0003)   &    (0.0003)   \\
    Missing number of tenants                &               &     -0.0899   &     -0.0786   &     -0.0795   \\
                                             &               &    (0.0803)   &    (0.0808)   &    (0.0810)   \\
    Number of tenants                        &               &      0.0004   &      0.0009   &     -0.0037   \\
                                             &               &    (0.0160)   &    (0.0162)   &    (0.0162)   \\
    Days since first published               &               &     -0.0003** &     -0.0003** &     -0.0002** \\
                                             &               &    (0.0001)   &    (0.0001)   &    (0.0001)   \\
    User has gender preferred by landlord    &               &      0.4865***&      0.4906***&      0.4818***\\
                                             &               &    (0.0571)   &    (0.0572)   &    (0.0574)   \\
    User has age preferred by landlord       &               &     -0.0068   &      0.0027   &      0.0095   \\
                                             &               &    (0.0536)   &    (0.0538)   &    (0.0539)   \\
    User has occupation preferred by landlord&               &      0.0771   &      0.0769   &      0.0799   \\
                                             &               &    (0.0851)   &    (0.0853)   &    (0.0855)   \\
    Has doorman                              &               &               &     -0.0924*  &     -0.1119** \\
                                             &               &               &    (0.0551)   &    (0.0554)   \\
    Has dishwasher                           &               &               &      0.0728   &      0.0729   \\
                                             &               &               &    (0.0492)   &    (0.0493)   \\
    Has terrace                              &               &               &     -0.0191   &     -0.0136   \\
                                             &               &               &    (0.0449)   &    (0.0450)   \\
    Has AC                                   &               &               &      0.0082   &      0.0149   \\
                                             &               &               &    (0.0471)   &    (0.0473)   \\
    Is smoker friendly                       &               &               &      0.0017   &     -0.0019   \\
                                             &               &               &    (0.0403)   &    (0.0404)   \\
    Has elevator                             &               &               &      0.0529   &      0.0416   \\
                                             &               &               &    (0.0413)   &    (0.0419)   \\
    Has exterior view                        &               &               &     -0.0260   &     -0.0161   \\
                                             &               &               &    (0.0399)   &    (0.0401)   \\
    Has TV                                   &               &               &     -0.0115   &     -0.0125   \\
                                             &               &               &    (0.0374)   &    (0.0374)   \\
    Has balcony                              &               &               &      0.1149***&      0.1049***\\
                                             &               &               &    (0.0381)   &    (0.0382)   \\
    Has heating                              &               &               &      0.0892** &      0.0945** \\
                                             &               &               &    (0.0395)   &    (0.0397)   \\
    In Sarri\`a Saint-Gervasi or Gr\`acia               &               &               &               &     -0.0710   \\
                                             &               &               &               &    (0.1956)   \\
    In Ciutat Vella or Saint Mart\'i           &               &               &               &     -0.1014   \\
                                             &               &               &               &    (0.1916)   \\
    In L’Eixample                              &               &               &               &     -0.0574   \\
                                             &               &               &               &    (0.1905)   \\
    In North of Barcelona                    &               &               &               &     -0.5561***\\
                                             &               &               &               &    (0.1947)   \\
    In Sants-Montjuic or Les Corts           &               &               &               &     -0.1696   \\
                                             &               &               &               &    (0.1869)   \\
    \hline
    Num Searches                             &      52,737   &      52,737   &      52,737   &      52,737   \\
    Num Results                              &      89,624   &      89,624   &      89,624   &      89,624   \\
    Pseudo-R2                                &      0.0005   &      0.0087   &      0.0111   &      0.0150   \\
    Log-likelihood                           &    -5112.57   &    -5070.45   &    -5058.32   &    -5038.15   \\
    \hline\hline
    \multicolumn{5}{p{14cm}}{\scriptsize\textit{Notes:} Table reports rank-ordered logit parameter estimates. Estimation via maximum likelihood. Standard errors in parentheses; *$p<0.10$, **$p<0.05$, ***$p<0.01$. The category ``In Greater Barcelona'' is omitted.}
    \end{tabular}
    \end{table}

    Next we discuss our parameter estimates of the click propensity. As \autoref{tab:click_bs} shows, we estimate a positive and statistically significant effect of the expected utility of a room, conditional on the available information, on the propensity to click on a room. This shows that users are more likely to click on rooms they are more likely to request. However, the estimated effect of the position a room is shown as a search result on the propensity that a user clicks on it is negative and statistically significant. Moreover, since the difference in click propensity from being shown at a higher position is higher for rooms with a higher utility, this effect is amplified for more preferred rooms.\footnote{Please bare in mind that ``being shown at a higher position'' is equivalent to a \textit{lower} value for the variable \textit{Position}: the top position has $Position=1$, etc.} Finally, being shown at the first position has an additional positive and statistically significant effect on the click propensity. As with our estimates for the request utility, in order to assess the magnitudes of these coefficients, we can normalize them for them to be measured in euros. \autoref{tab:norm_click_bs} in \autoref{appendix:more_tables} reports the normalized values of the coefficients in \autoref{tab:click_bs}.\footnote{Normalizing the coefficients of the click propensity is done by dividing them by the absolute value of the utility coefficient multiplied by the absolute value of the price coefficient in the request utility. See that this is necessary in order to obtain the marginal effect of a change in the expected utility when it is measured in euros.} Accordingly, the negative effect that being shown in a lower position has on the click propensity can be valued at around \euro 200--250 \textit{per position}. And being shown at the top position has an additional positive effect equivalent to \euro 352.

    \begin{table}[t]
    \centering\small
    \caption{Parameter estimates of click propensity\label{tab:click_bs}}
    \begin{tabular}{lccccc}\hline\hline
                             &         (1)   &         (2)   &         (3)   &         (4)   \\\hline
    Position                 &     -0.0538***&     -0.0834***&     -0.0698***&     -0.0698***\\
                             &    (0.0008)   &    (0.0026)   &    (0.0049)   &    (0.0049)   \\
    Position$^2$             &               &      0.0016***&      0.0010***&      0.0010***\\
                             &               &    (0.0001)   &    (0.0002)   &    (0.0002)   \\
    1(Position = 1)          &               &               &      0.0991***&      0.0987***\\
                             &               &               &    (0.0212)   &    (0.0212)   \\
    1(Position = 2)          &               &               &     -0.0014   &     -0.0010   \\
                             &               &               &    (0.0182)   &    (0.0182)   \\
    1(Position = 3)          &               &               &     -0.0028   &     -0.0027   \\
                             &               &               &    (0.0158)   &    (0.0158)   \\
    $\mathbb{E}[u|I]$        &               &               &               &      0.3027***\\
                             &               &               &               &    (0.0448)   \\
    $\mathbb{E}[u|I]$ $\times$ Position&               &               &               &      0.0320***\\
                             &               &               &               &    (0.0044)   \\\hline
    Num Searches             &     173,223   &     173,223   &     173,223   &     173,223   \\
    Num Results              &   2,066,147   &   2,066,147   &   2,066,147   &   2,066,147   \\
    Pseudo-R2                &      0.0143   &      0.0147   &      0.0149   &      0.0161   \\
    Log-likelihood           &  -1.644e+05   &  -1.644e+05   &  -1.643e+05   &  -1.641e+05   \\
    \hline\hline
    \multicolumn{5}{p{11cm}}{\scriptsize\textit{Notes:} Table reports rank-ordered logit parameter estimates. Estimation via maximum likelihood. Standard errors in parentheses (do not account for variation in first-stage parameters); *$p<0.10$, **$p<0.05$, ***$p<0.01$.}
    \end{tabular}
    \end{table}

\begin{figure}[htbp!]
        \centering
        \subfloat[Click index\label{fig:5-top}]{\includegraphics[width=0.75\textwidth]{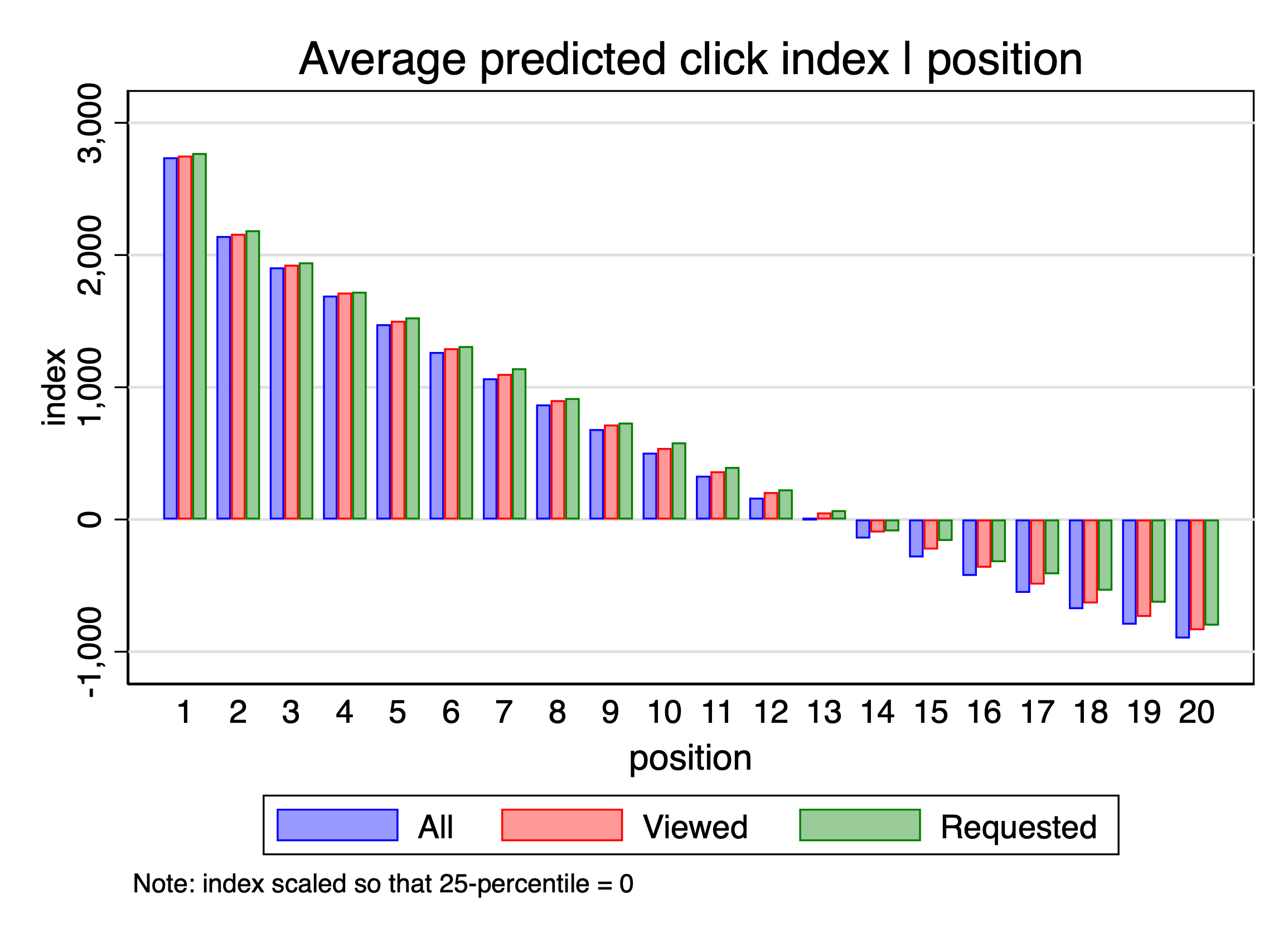}}

        \vspace{0.5cm}
        
        \subfloat[Request Utility]{\includegraphics[width=0.75\textwidth\label{fig:5-bottom}]{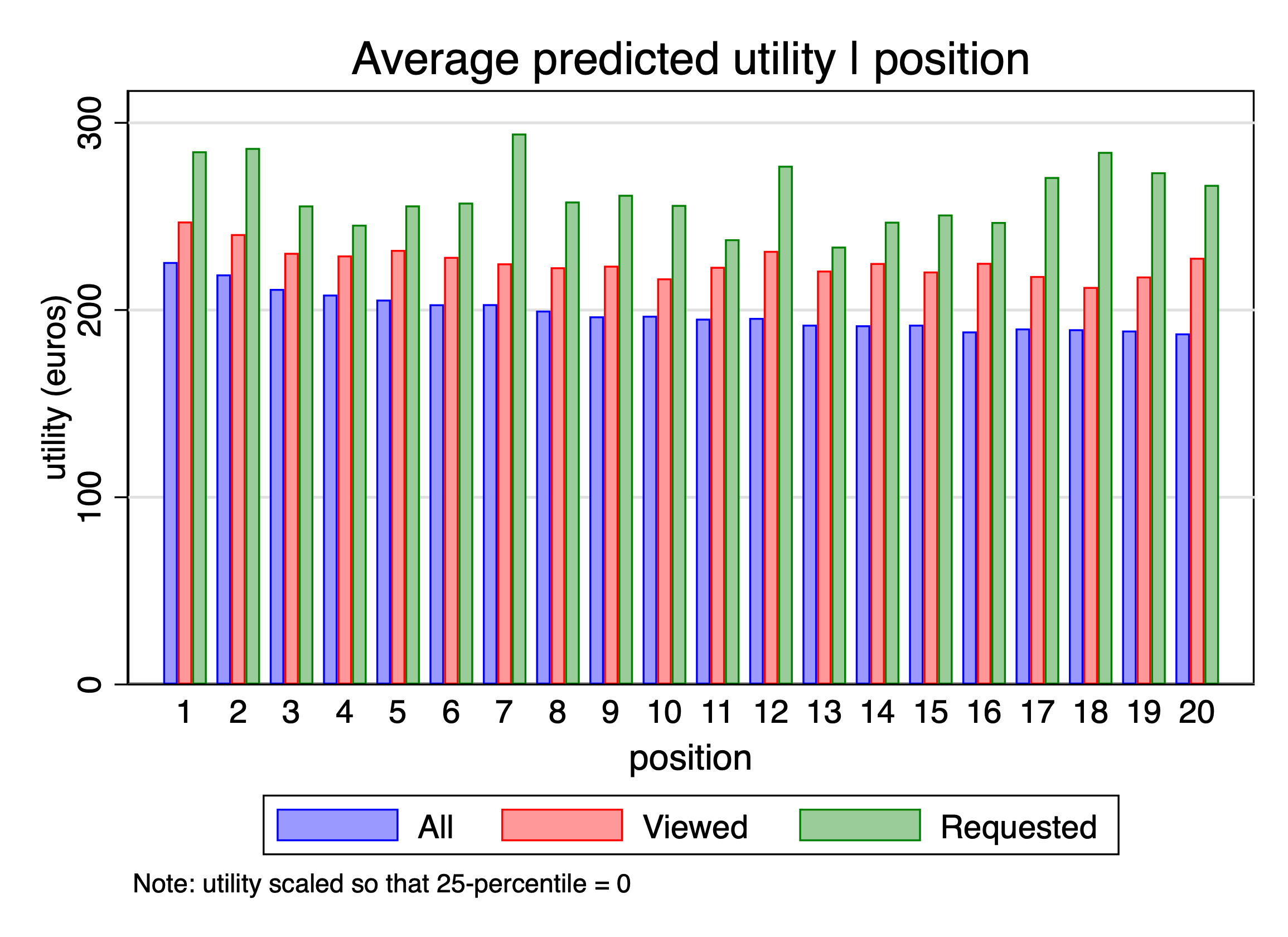}}
        \caption{Estimated click index and request utility\label{fig:5}}
    \end{figure}

    Finally, as we mentioned above, in the counterfactuals below, we will use the estimated utility and click propensity indices. \autoref{fig:5} plots these two indices against the position in which rooms are shown.\footnote{In the plots, we normalize the location of the indices, so that the room at the 25th percentile has a zero index, both for the click propensity and the request utility.} The top graph (\autoref{fig:5-top}) shows that the propensity to click on a room decreases monotonically with the position a room is shown. And this holds true in all rooms in the data, regardless of whether they are clicked on or requested. The bottom graph (\autoref{fig:5-bottom}) shows that the predicted request utility of rooms is relatively flat and does not depend heavily on the position they are shown. This can be seen especially in the predicted utility of the rooms that are requested in the data (green bars in \autoref{fig:5-bottom}). Altogether this shows that our estimates of the click propensity and request utility echo the raw data (cf.\ \autoref{fig:2}): while users are more likely to click at the top, once they have clicked on a room, the position a room is shown in has no effect on whether the user requests that room.
    
\section{Counterfactuals}\label{sec:counterfactuals}

Based on the estimation results above, we simulate a series of counterfactuals in which we consider the effect of changing the search ranking algorithm on the click and request outcomes. The counterfactual exercises below consider the following question: What would the users have clicked on and requested had they observed the same set of search results but ranked in a different order? Throughout we assume that changing the ranking algorithm would not affect the users' unmodeled search behavior: the number of searches they conduct, clicks they make, and requests they submit.

Before presenting the results, we note that in these simulations we do not incorporate the utility shocks when predicting outcomes; we only make use of the estimated indices to predict clicks and requests. This approach is more robust because it does not rely on the parametric assumptions on the utility shocks, but solely on the heterogeneity present in the room and user characteristics. 
In this sense, our approach is conservative at capturing horizontal differentiation in preferences.\footnote{In \autoref{appendix:clusters}, we investigate a more flexible specification that allows to capture additional heterogeneity present in the search data.}
    
\subsection{Different ranking algorithms: effects of congestion}
      
\begin{figure}[t]
        \centering
        \subfloat[Predicted clicks\label{fig:6a}]{\includegraphics[width=0.5\textwidth]{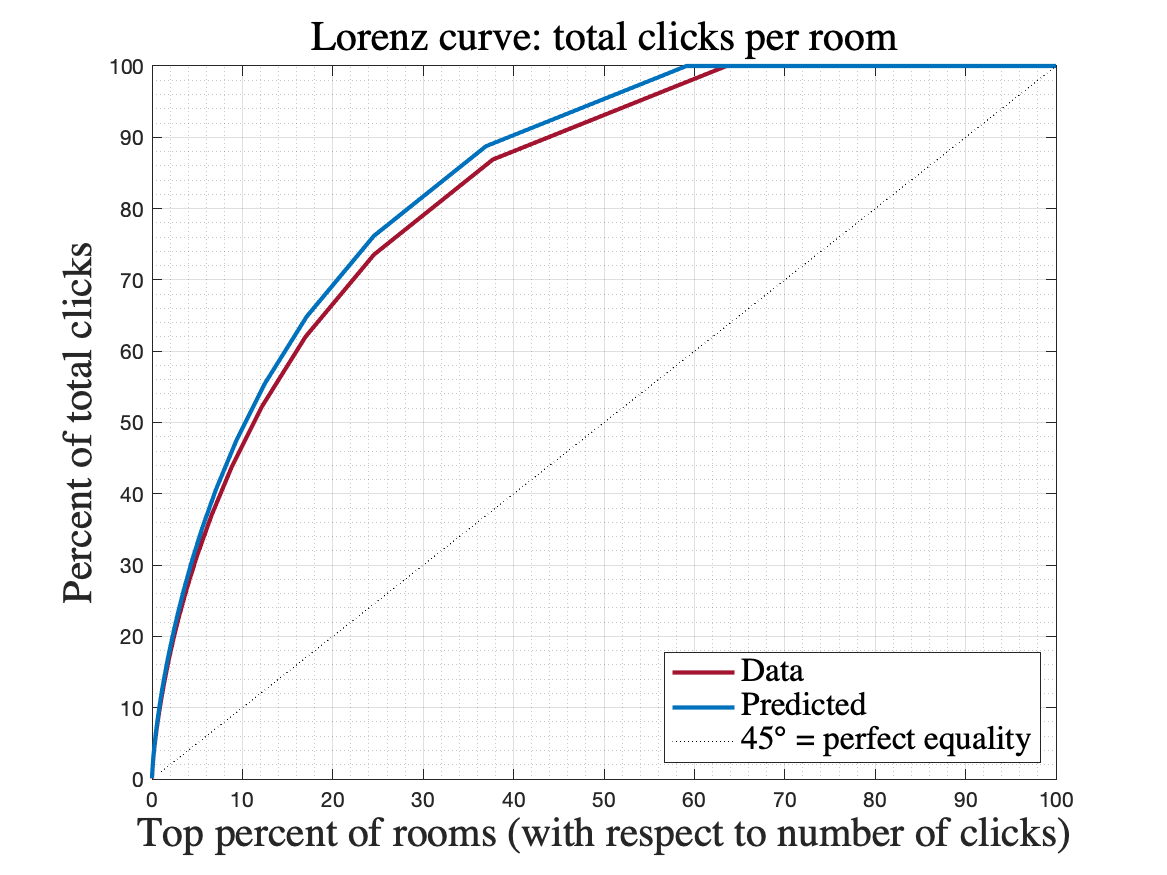}}
        \subfloat[Predicted requests\label{fig:6b}]{\includegraphics[width=0.5\textwidth]{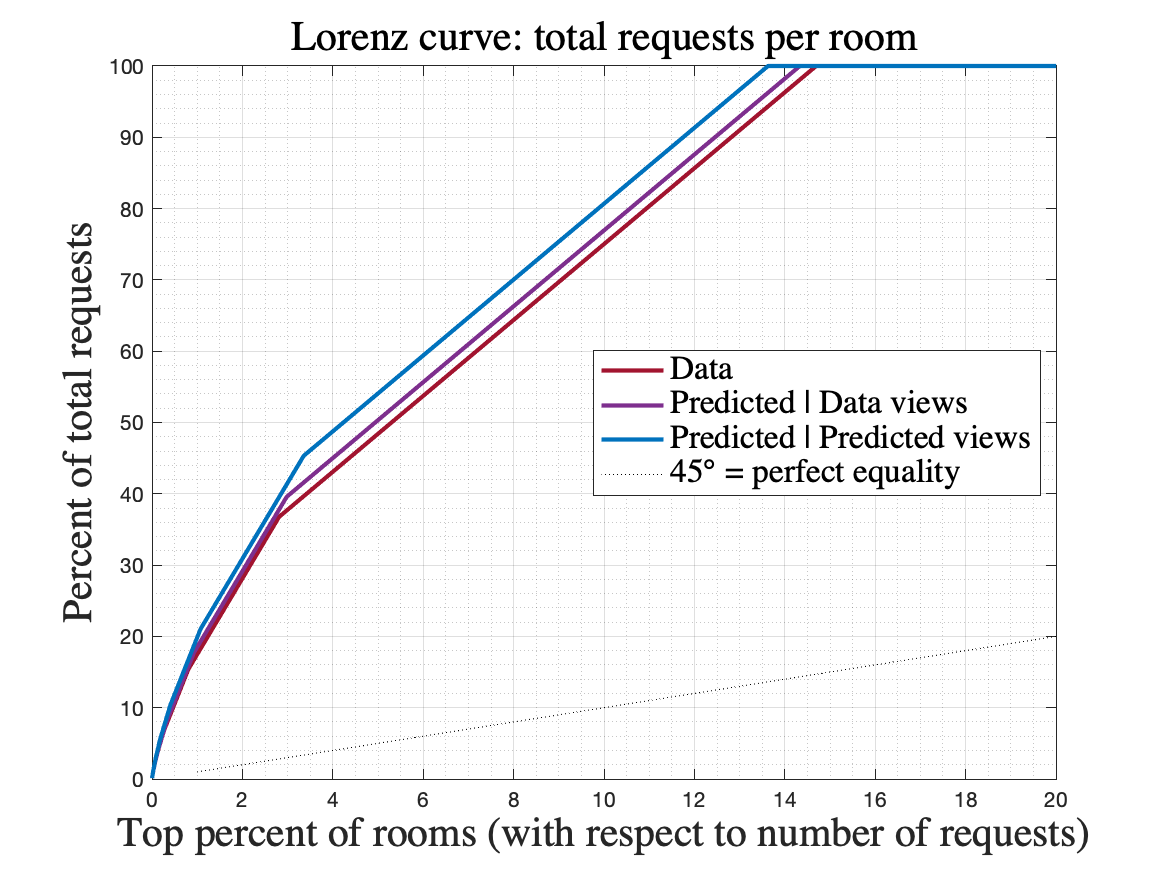}}

        \subfloat[Counterfactual clicks\label{fig:6c}]{\includegraphics[width=0.5\textwidth]{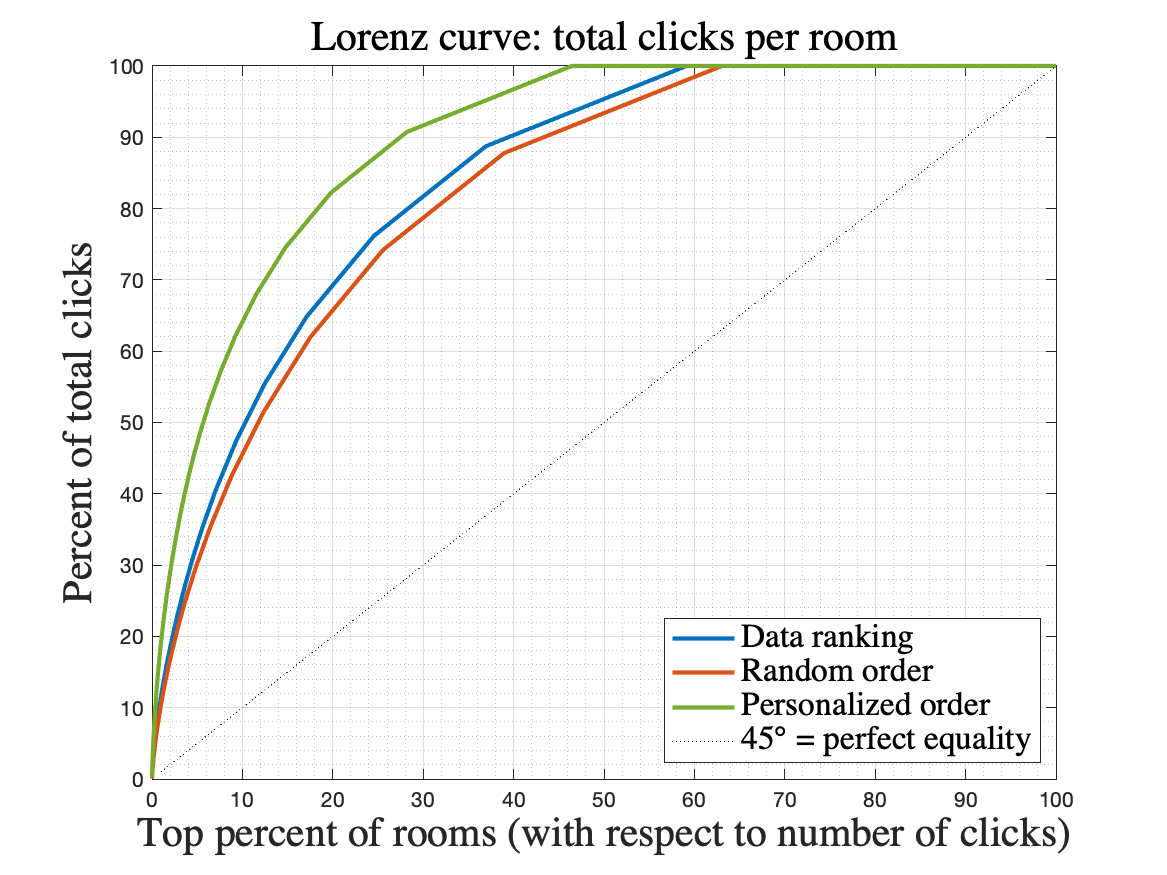}}
        \subfloat[Counterfactual requests\label{fig:6d}]{\includegraphics[width=0.5\textwidth]{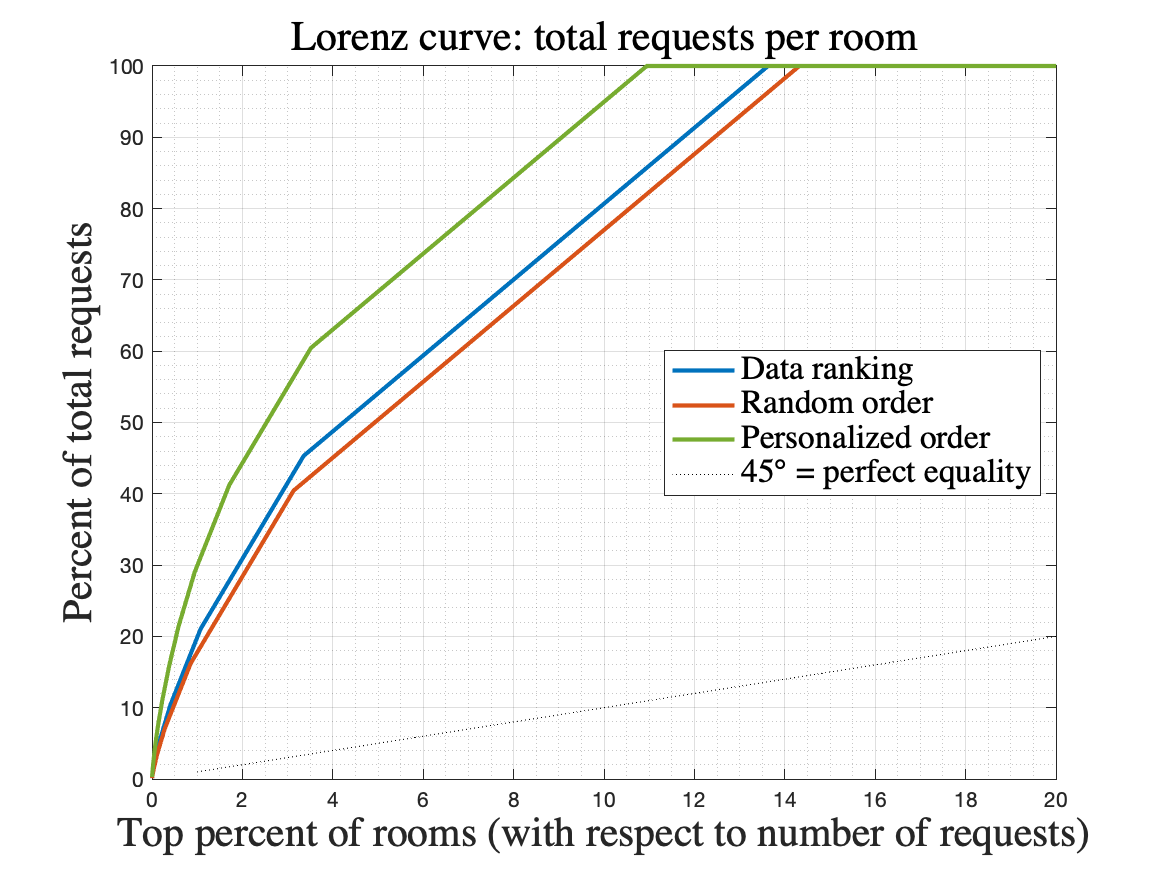}}
          
        \caption{Lorenz curves of the distribution across rooms of clicks and requests\label{fig:6}}
    \end{figure}

    In the counterfactuals we explore alternative ranking algorithms and compare their performance to the platform's current ranking system, which surfaces the same ordering of search results to every user. To compare congestion across counterfactual scenarios, we again use Lorenz curves, as in \autoref{fig:3} above. \autoref{fig:6} presents Lorenz curves across a number of scenarios, which we discuss in sequence. The top two graphs in \autoref{fig:6} show the Lorenz curves corresponding to the data and predicted values derived from simulations of the model at the estimated model parameters. The close relation between the data (red curve) and predicted values (blue curve) confirm that the estimated model fits the data well, both in terms of the clicks (\autoref{fig:6a}) and requests (\autoref{fig:6b}).

In the bottom two graphs of \autoref{fig:6}, we consider two counterfactual scenarios that incorporate our parameter estimates of user preferences, but only change the order in which search results are shown to users, while keeping the selection of results within each search fixed.
    
First, we consider the case of {\bfseries full personalization} in which the search results within each search are ordered according to each user's preferences. Since users tend to click on rooms shown at top positions, this algorithm steers users towards the rooms they prefer the most. The Lorenz curve from this full personalization scenario is graphed in green.  Second, we consider the case of {\bfseries random order} in which the order in which rooms are shown is fully randomized within every search. This algorithm destroys the correlation between the position at which rooms are shown across searches and users, and is aimed at reducing congestion. This scenario is shown in the orange-colored Lorenz curves. 

The bottom two graphs show that these counterfactual ranking algorithms do change the concentration of clicks (\autoref{fig:6c}) and requests (\autoref{fig:6d}). The direction of the change in clicks and requests is qualitatively similar: randomizing the order in which rooms are shown (orange line) decreases congestion, if only slightly, while the fully personalized one (green) increases congestion substantially. That is, using the information on user preferences to rank search results within searches leads to higher congestion compared to the data. 

This suggests that across users, the rooms are primarily vertically differentiated, so that most users' preferences agree on which rooms offer the highest utility. This echoes the findings of \cite{ChenHsiehLin2021}, who likewise found that surfacing dating prospects to users in order of their reported preferences results in excess congestion.  From a congestion perspective, then, the homogeneous ranking algorithm currently used by the platform emerges as a preferable option to a fully personalized one. 

\subsection{Utility and congestion trade-off}\label{sec:tradeoff}

    The results above illustrate how congestion in this marketplace can be reduced by randomizing search results.  However, random search results imposes a potential cost of lowering the utility that users receive from the rooms that they click on and request.
At the same time, personalizing search results to users by showing them rooms in the order of their preferences can optimize utility, but result in greater congestion.

This suggests a key trade-off in this market, between congestion and utility.  To examine this trade-off more formally, we introduce a continuum of ranking algorithms indexed by the parameter $\alpha\in [0,1]$, where $\alpha$ denotes the weight given to the ``fully personalized'' order, which ranks more preferred rooms at the top, and $1-\alpha$ the weight given to the ``random order,'' which ranks them randomly. In \autoref{fig:6} above, we only considered the polar cases of $\alpha = 1$ (``full personalization''), and $\alpha = 0$ (``random order''); here we expand to consider all intermediate alternatives. 

\begin{figure}[t]
        \centering
        \subfloat[$\alpha$-plot clicks\label{fig:7a}]{\includegraphics[width=0.5\textwidth]{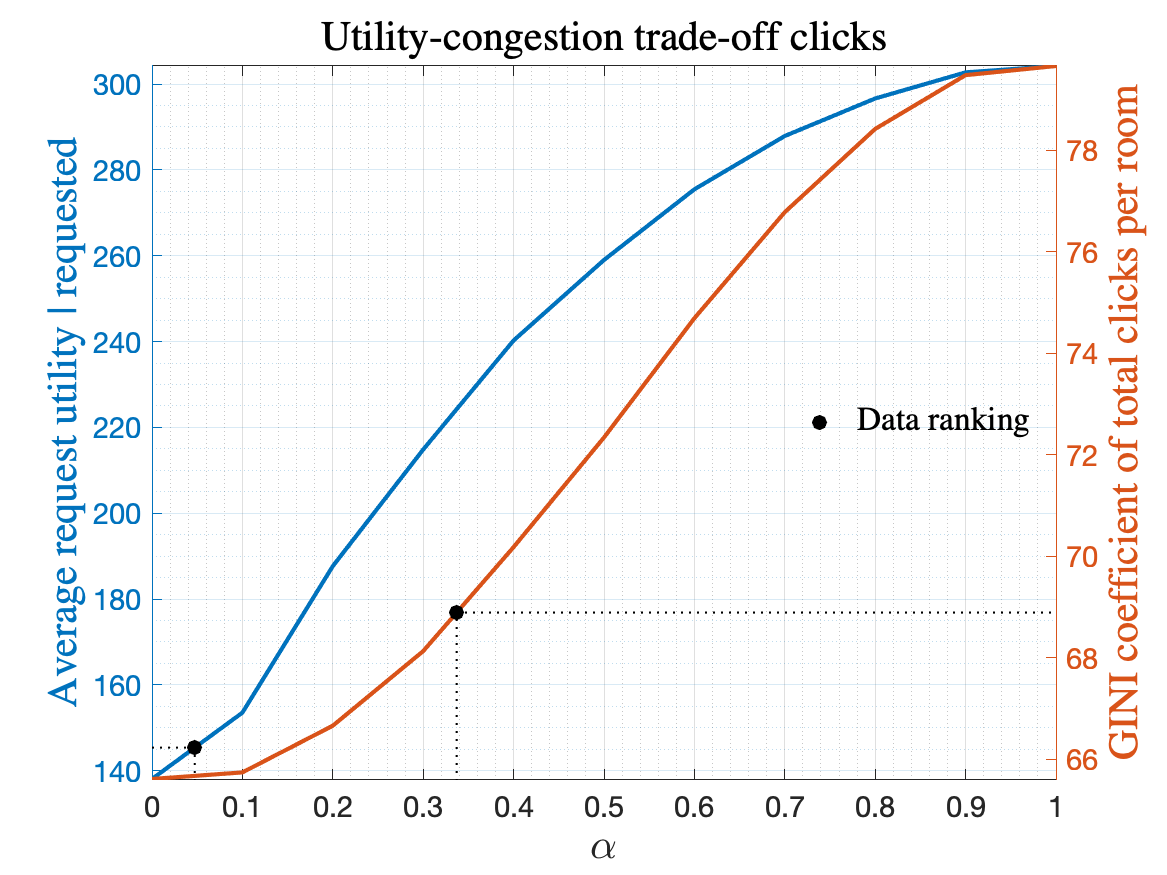}}
        \subfloat[Efficiency frontier for clicks\label{fig:7b}]{\includegraphics[width=0.5\textwidth]{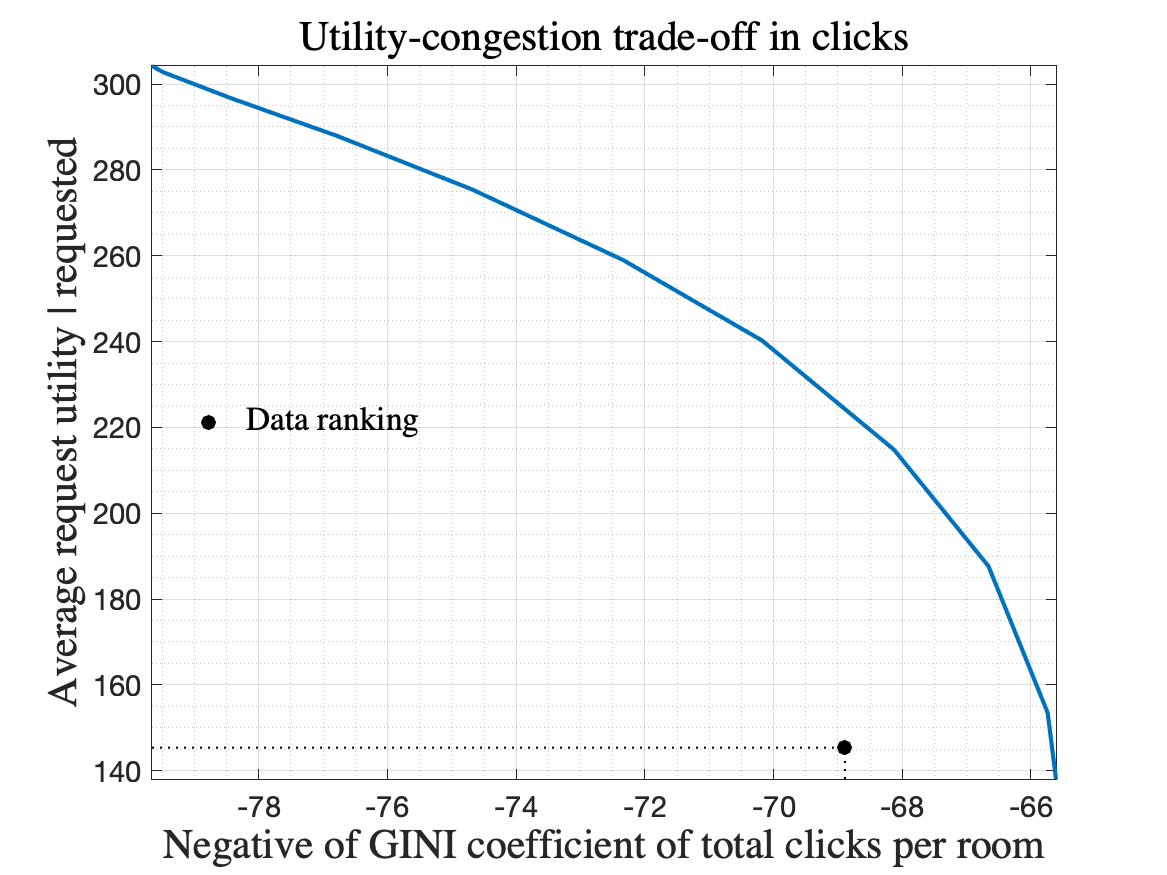}}
        
        \subfloat[$\alpha$-plot requests\label{fig:7c}]{\includegraphics[width=0.5\textwidth]{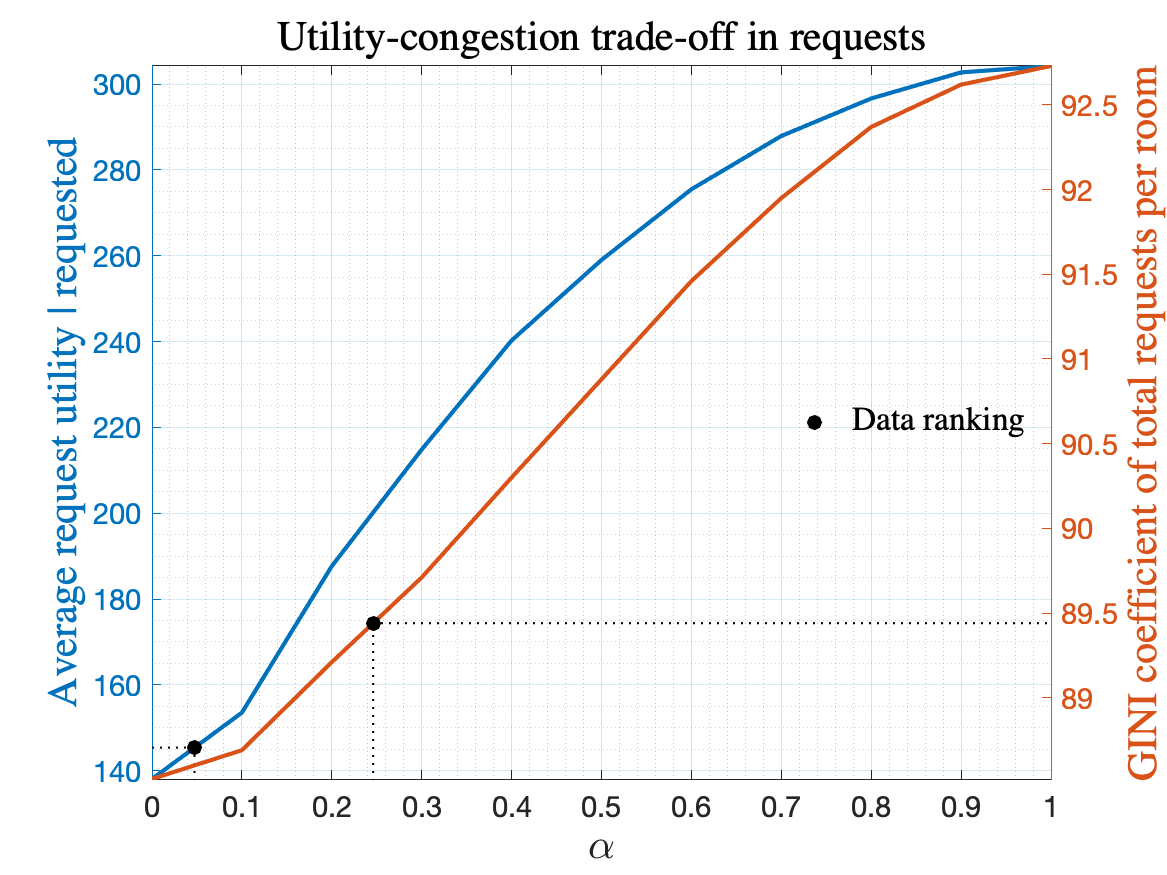}}
        \subfloat[Efficiency frontier for requests\label{fig:7d}]{\includegraphics[width=0.5\textwidth]{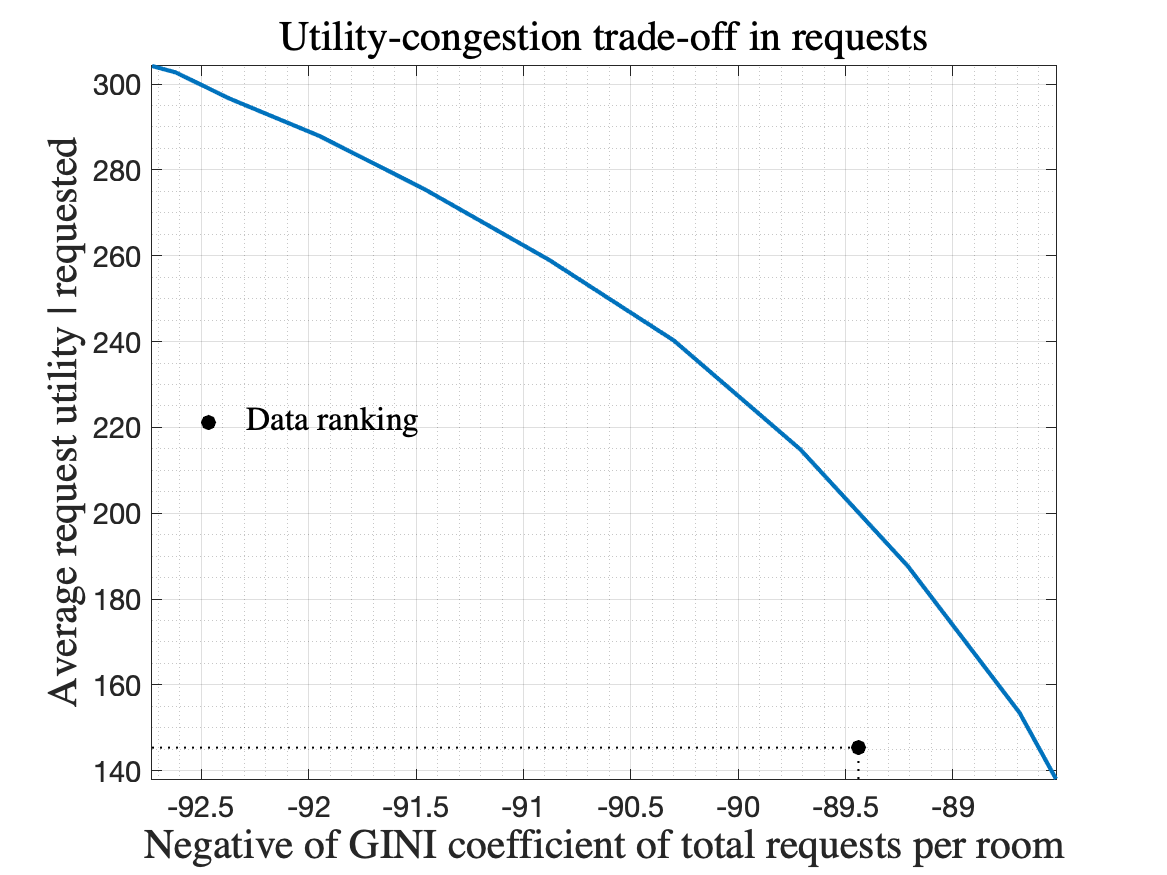}}
        \caption{Utility and congestion trade-off of clicks and requests\label{fig:7}}
    \end{figure}

The effects of incorporating information on user preferences to a random ranking algorithm, i.e., personalizing it, is illustrated in \autoref{fig:7}.  In the left two graphs, the $x$-axis plots $\alpha$ moving from 0 to 1.  The blue curve shows how the average utility of the rooms requested by users increases with greater personalization, as $\alpha$ goes from 0 to 1. The red curve shows how congestion also increases with greater personalization. This pattern is present both in the distribution of clicks (\autoref{fig:7a}) and requests (\autoref{fig:7c}). 

The utility and congestion levels in the data are indicated by black dots in the plots. In the left two graphs of \autoref{fig:7}, one can see that the average utility in the data can be achieved by incorporating little information about user preferences to an otherwise random ranking algorithm, i.e., by setting a value for $\alpha$ lower than 0.1. Furthermore, the graphs also show that setting such value of $\alpha$ would result in lower congestion than what is observed in the data since the congestion observed in the data corresponds to higher values of $\alpha$, around 0.34 for clicks and 0.25 for requests. 

The above leads to two conclusions: (i) the slopes of the curves describing the change in average utility and congestion encode the trade-off between utility and congestion in this market, and (ii) the utility and congestion levels observed in the data seem to be inefficient, in that we can improve one without worsening the other. These conclusions can be appreciated in the right two graphs of \autoref{fig:7}, which plot possibility frontiers in the (utility, congestion) space achievable by different values of $\alpha$.

The black dot in these plots indicates the (utility, congestion) point observed in the data.  Notably, the black dot lies {\em below} the frontier of both clicks (\autoref{fig:7b}) and requests (\autoref{fig:7d}), illustrating how the current ranking algorithm utilized by the platform is inefficient: holding congestion fixed, it is possible to improve utility by incorporating more personalization; holding utility fixed, it is possible to lower congestion by introducing more randomization.

\subsection{Looking beyond the data: a hypothetical ``garbling'' scenario}

\begin{figure}[t]
        \centering
        \subfloat[$\alpha$-plot clicks]{\includegraphics[width=0.5\textwidth]{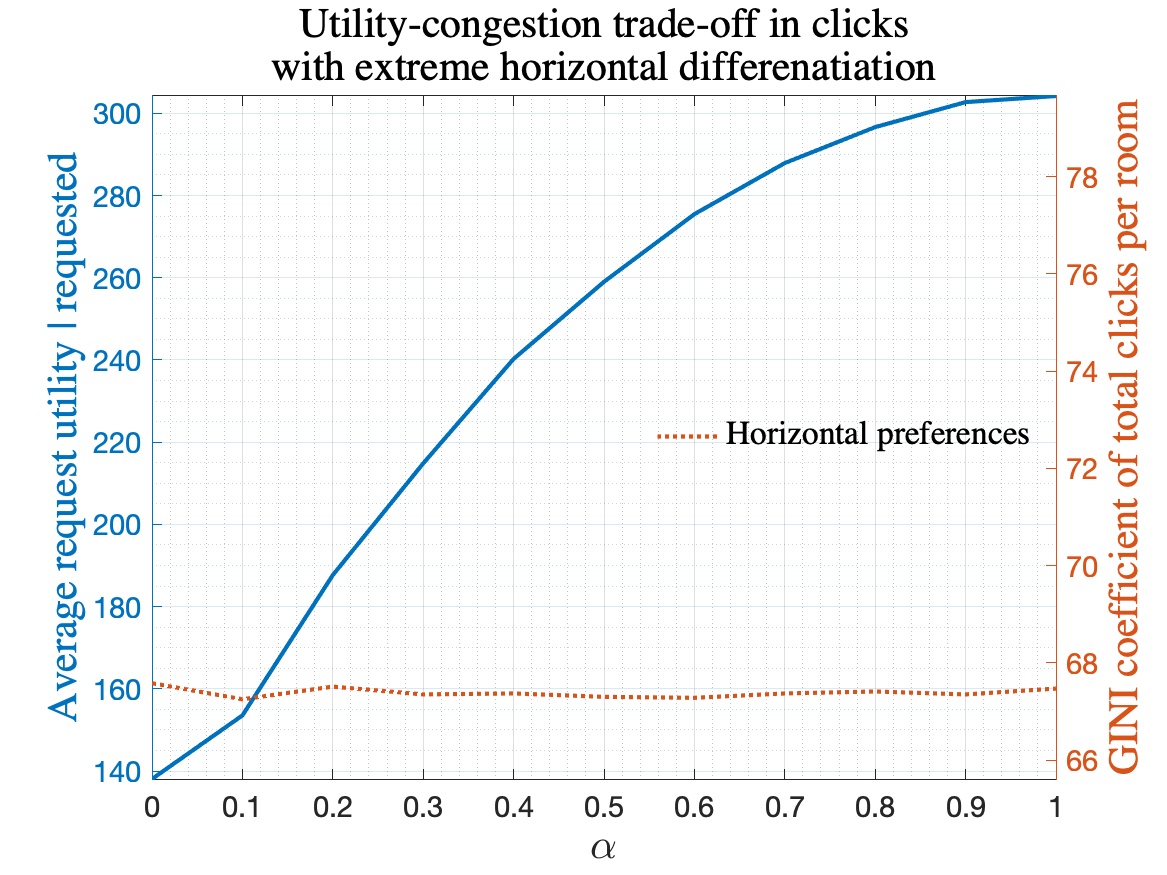}}
        \subfloat[Efficiency frontier for clicks]{\includegraphics[width=0.5\textwidth]{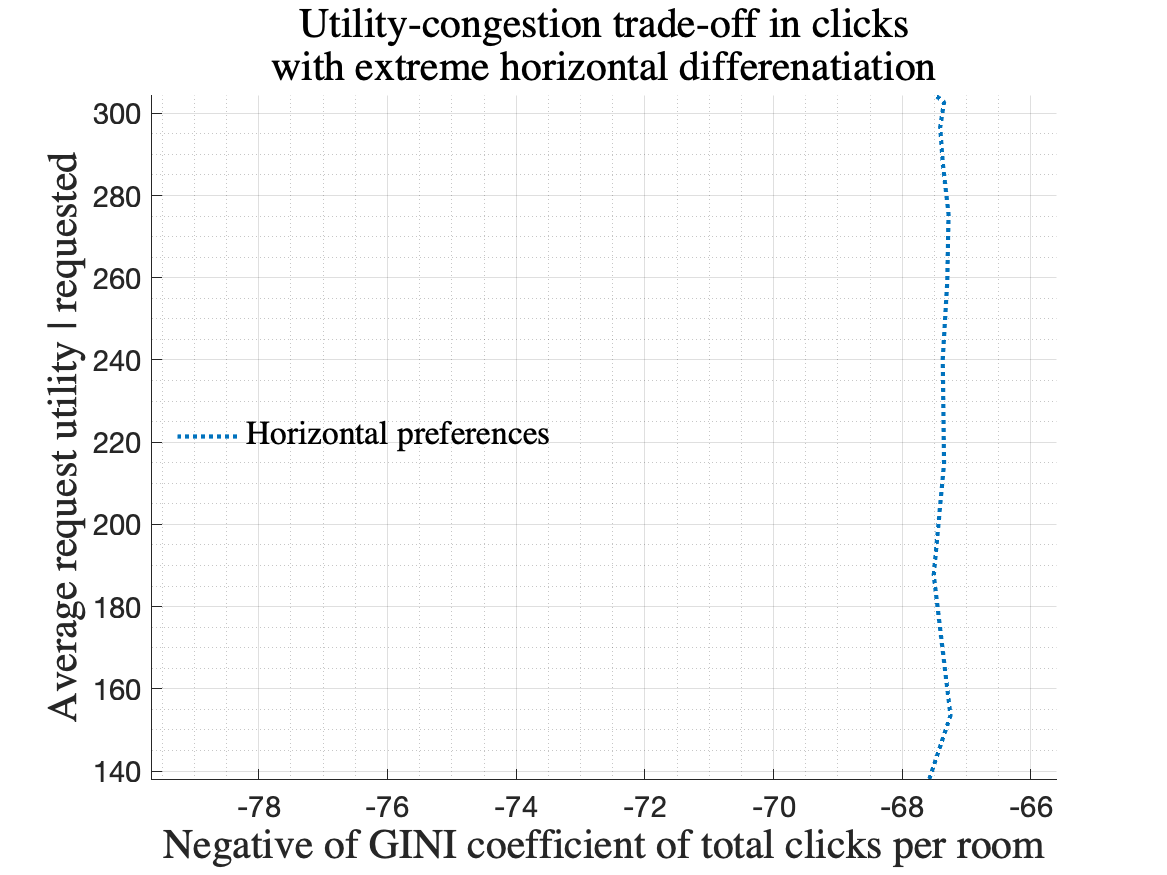}}
        
        \subfloat[$\alpha$-plot requests]{\includegraphics[width=0.5\textwidth]{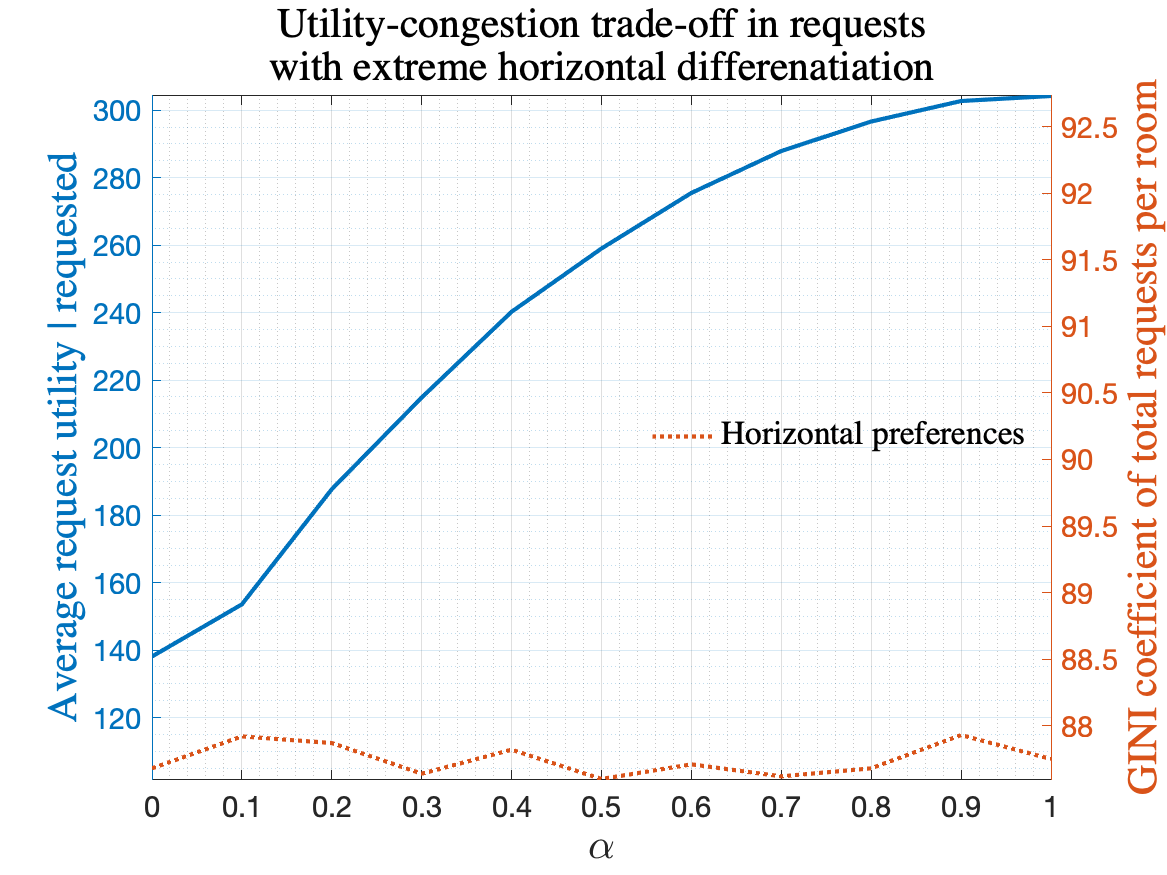}}
        \subfloat[Efficiency frontier for requests]{\includegraphics[width=0.5\textwidth]{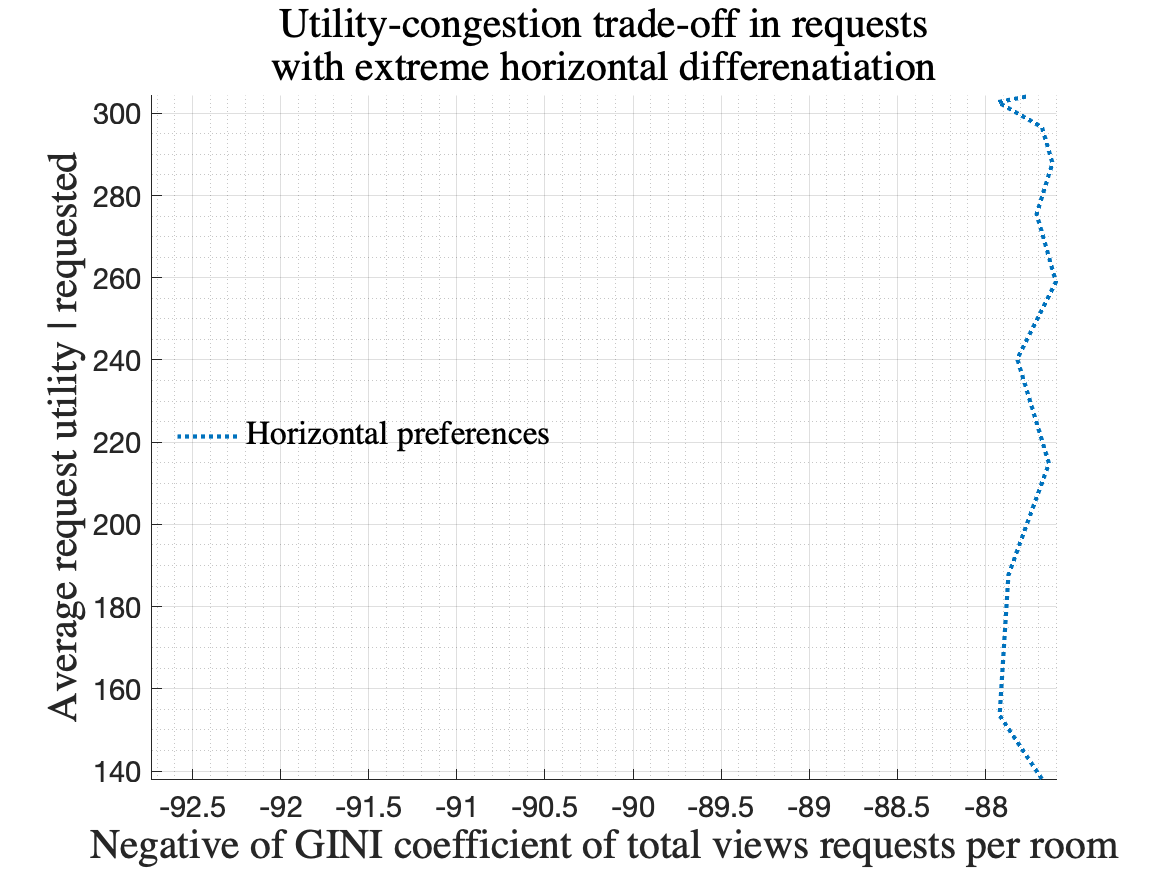}}
        \caption{Utility and congestion trade-off of clicks and requests \\with extreme horizontal differentiation\label{fig:8}}
    \end{figure}

The trade-off between utility and congestion depends on the degree of differentiation in consumer preferences. To emphasize this, in Figure \ref{fig:8}, we illustrate a tantalizing hypothetical scenario in which users' preferences are extremely horizontally differentiated.  We achieve this by randomly relabeling the room IDs across searches. While this ``garbling'' keeps the utility of the alternatives in each search unchanged, including the utility of requested rooms, it destroys the correlation of preferences across users. In such a case, we see that congestion no longer increases in $\alpha$, which eliminates the trade-off between utility and congestion (see the dotted red line in the left two graphs of \autoref{fig:8}). And the corresponding efficiency frontier approximates a vertical line (dotted blue line in right two graphs of \autoref{fig:8}), indicating that at a fixed level of congestion one can improve utility arbitrarily by personalizing search results to users. 

Intuitively, when users' preferences are extremely horizontally differentiated, a ``free lunch'' emerges: high degrees of personalization, which lead to high user utility, do not increase congestion because user preferences are highly dissimilar. The nature of user preferences plays a central role in the direction of congestion. This exercise points out the potential benefits that firms may gain if they are able to ``horizontalize'' user preferences, perhaps by emphasizing features of products along which user preferences are more likely to disagree. 
  We will study these issues in future work.

\section{Conclusions}\label{sec:conclusion}

Two-sided platforms have the opportunity to reduce search costs of match seekers. However, unnecessary congestion may be generated if users are steered towards the same alternatives, for example, by showing alternatives in the same order to users. This may be a concern especially if the platform has limited information on user preferences. Using detailed data from a room-rental platform we estimate preferences over rooms and demonstrate
    that, for this market, there is a key trade-off between congestion and utility, in that ranking algorithms which reduce congestion must also reduce personalization by steering users to lower-utility rooms.
    
    This arises from the fact that estimated preferences are primarily vertically differentiated, so that customers tend to agree on the most desirable features in the available rooms.
    Indeed, in hypothetical simulations, we show that if preferences exhibited sufficient horizontal differentiation, then the trade-off would reverse, leading to a "free lunch" whereby personalized algorithms which present customers their most desired rooms would not reduce congestion.
    
    Clearly, platforms cannot change customers' preferences.  However, they do have a role to play in designing their websites to emphasize different room features at different stages of the customer search journey. Can platforms effectively ``horizontalize'' their users' preferences by emphasizing the horizontal rather than vertical features of available rooms?  These and related questions will be explored in future work.

\clearpage
\bibliographystyle{ecta}
\bibliography{mybib}
\clearpage
\appendix
\section{Appendix}\label{sec:appendix}
\subsection{Search filters and user heterogeneity\protect\footnote{Some of the results in this section are preliminary and still need to be updated.}\label{appendix:clusters}}

In this section, we consider a more flexible version of our model that incorporates unobserved heterogeneity present in the search data. As explained above, the goal of this portion of the analysis is to show how the utility congestion trade-off we estimate becomes steeper with a more flexible model that captures more heterogeneity.

The intuition behind our approach is that users who have different preferences might search differently, for example, by specifying different search criteria when using search filters. In our platform, a number of search filters are available for price range, location (neighborhood) of the room, amenities, gender-preferences, etc. While we have no direct information on the filters users employ when searching, we can assume that a user has filtered for a specific characteristic if all the results within a search share that characteristic. Then, if we observe a user that filters searches frequently by a given characteristic (e.g., students search for rooms near the university, females search for women-only rooms, couples look for rooms without other tenants, etc.), this may indicate that their preferences are different to those of users who tend \textit{not} to filter according to the same characteristics (e.g., professionals search for rooms near the city center, men do not search for women-only rooms, singles search for rooms with roommates, etc.).

Our approach captures the heterogeneity present in search data by clustering users according to the frequency with which they filter by each characteristic. Subsequently, we estimate request and click choice models, same as we did above, for each cluster separately, and in this way allow the estimated parameters to vary across clusters.

In order to partition our sample of users in clusters, we use the $k$-medoids clustering method over the percentage of searches a user makes that are filtered by the different room characteristics. For example, if students filter more often for rooms near the university, our method would assign them to a different cluster as users who do not filter for rooms close the university. We use the percentage of filtered searches to accommodate the wide variation in the search behavior of users.

The $k$-medoids clustering method is similar to the $k$-means clustering method to discretize unobserved heterogeneity \citep{Bonhomme2022}. Both methods split the observations in a dataset into $k$ clusters by minimizing the distance between the data points to the center of their cluster. In $k$-means, the center of a cluster is the average between the points in a cluster and is referred to as a \textit{centroid}. In $k$-medoids the center of a cluster is an actual point within the cluster, and is referred to as a \textit{medoid}. The difference between both methods is akin to that of using means or medians to describe data. $k$-medoids is a more robust method for outliers, and particularly useful for categorical data or cases in which the centroids of the clusters are not data points. In particular, the medoids of the clusters found by $k$-medoids are simpler to interpret than the centroids found by $k$-means.\footnote{For a thorough discussion on the difference between both methods and how to implement them, see \cite{Kaufman1990}.} We use $k$-medoids because many users never use filters for some characteristics. Even though the dataset we use to determine clusters is not composed of categorical data, it has many zeros. 

As discussed above, before estimation we pre-process the data to cluster the users in our sample to allow for user-specific preference heterogeneity.   In the preferred specification, we split the users into three clusters using \textit{k-medoids}.\footnote{As with $k$-means, the number of clusters in $k$-medoids needs to be manually prespecified. We tried with alternative number of clusters and found three to be the most sensible; in particular, using more clusters resulted in clusters with very few users.} \autoref{tab:5} reports the medoids of the clusters, and \autoref{tab:6} their size and summary statistics of the users in each cluster. Importantly, the characteristics in \autoref{tab:6} are \textit{not} used to compute the clusters, just the frequencies with which users filter searches by any of the characteristics reported in \autoref{tab:5}.

    \begin{table}[t]
        \centering
        \caption{Medoids of each cluster ($k$-medoids) \label{tab:5}}
        \begin{tabular}{lccc}\hline\hline
         & Cluster 1 & Cluster 2 & Cluster 3 \\\hline
        User has gender preferred by landlord & 12.98 & 20.83 & 68.54 \\
        User has age preferred by landlord & 4.21 & 4.17 & 15.05 \\
        User has occupation preferred by landlord & 79.30 & 4.17 & 76.33 \\
        Has doorman & 0.35 & 0.00 & 0.27 \\
        Has dishwasher & 0.35 & 4.17 & 0.55 \\
        Has terrace & 0.00 & 0.00 & 0.55 \\
        Has AC & 0.35 & 4.17 & 0.55 \\
        Is smoker friendly & 2.46 & 4.17 & 1.09 \\
        Has elevator & 0.70 & 4.17 & 8.21 \\
        Has exterior view & 3.51 & 8.33 & 3.15 \\
        Has TV & 4.91 & 4.17 & 3.69 \\
        Has balcony & 0.00 & 4.17 & 2.74 \\
        Has heating & 3.16 & 4.17 & 2.05 \\
        In Saint-Gervasi or Gracia & 0.00 & 0.00 & 2.74 \\
        In Ciutat Vella or Saint Marti & 3.86 & 0.00 & 7.66 \\
        In Eixample & 1.05 & 0.00 & 13.27 \\
        In North of Barcelona & 2.81 & 0.00 & 0.14 \\
        In Sants-Montjuic of Les Corts & 1.05 & 0.00 & 6.43 \\\hline\hline
        \multicolumn{4}{p{12cm}}{\scriptsize\textit{Notes:} Clusters are constructed over the percent of searches a user makes filtering by each covariate. A search is said to be filtered if all search results have the covariate = 1.}
        \end{tabular}
    \end{table}

    \begin{table}[htbp!]
        \centering
        \caption{Average user characteristics within each cluster \label{tab:6}}
        \begin{tabular}{lccc}\hline\hline
         & Cluster 1 & Cluster 2 & Cluster 3 \\\hline
        Percent of users ($\%$) & 23.33 & 35.40 & 41.27 \\
        Age & 27.53 & 29.19 & 30.64 \\
        Is female & 0.50 & 0.93 & 0.17 \\
        Is student & 0.59 & 0.31 & 0.24 \\
        Is worker & 0.31 & 0.93 & 0.99 \\
        Number of searches per user & 137.17 & 174.79 & 143.61 \\
        Number of clicks per user & 141.46 & 169.94 & 137.77 \\
        Number of requests per user & 12.58 & 13.83 & 13.77 \\\hline\hline
        \multicolumn{4}{p{11cm}}{\scriptsize\textit{Notes:} Clusters are constructed over the percent of searches a user makes filtering by each covariate. A search is said to be filtered if all its results have the covariate = 1. User characteristics are not used to construct the clusters.}
        \end{tabular}
    \end{table}

The three clusters have reasonable interpretations. As shown in \autoref{tab:6}, users in Cluster 3, which is the largest, are primarily workers ($>99$\%) and men ($>80$\%).   Women workers make up most of Cluster 2, while Cluster 1 contains many students ($\sim 60\%$) and is roughly split evenly between men and women. 

\autoref{tab:7} reports the estimated parameters of the request utility, and \autoref{tab:8} those of the click propensity. We allow the parameter vector to vary across clusters in both cases. As we do in the analysis in the main text, we use the parameter estimates to compute utility and click-propensity indices for every user. This allows us to predict clicks and requests, in particular, when the order in which rooms are shown is changed. We conduct the same analysis we do in the text varying $\alpha\in[0,1]$, the weight given to the fully personalized ranking in which every user is shown rooms ranked according to their estimated preferences, relative to random order.

\begin{figure}[htbp!]
        \centering
        \subfloat[$\alpha$-plot clicks]{\includegraphics[width=0.5\textwidth]{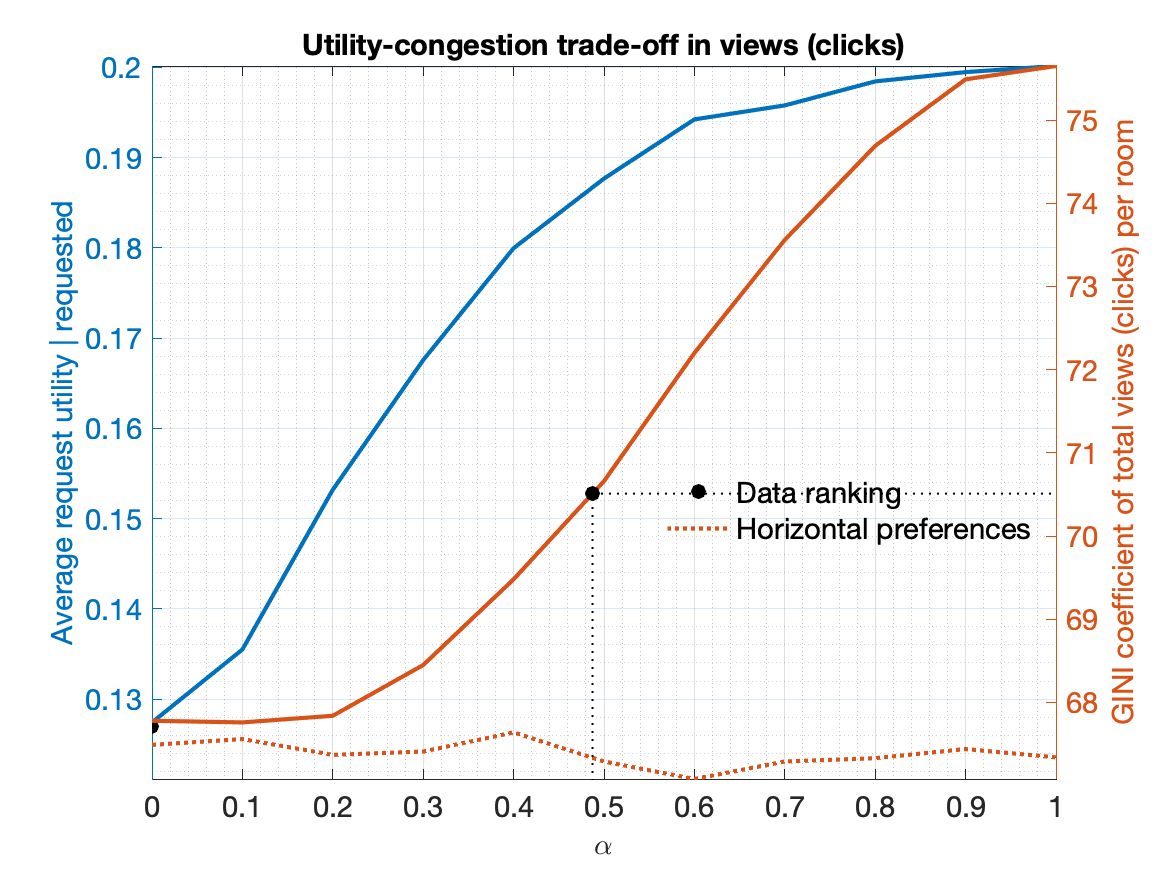}}
        \subfloat[Frontier clicks]{\includegraphics[width=0.5\textwidth]{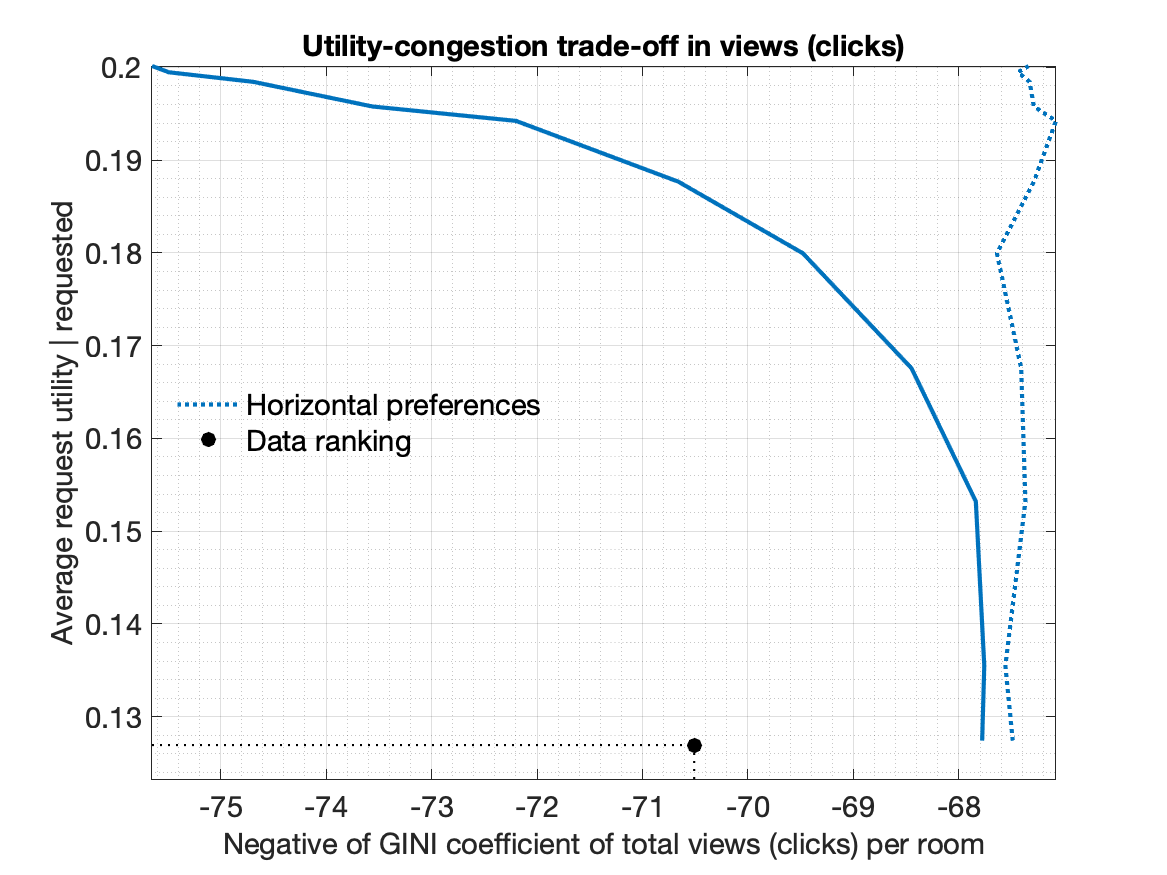}}
        
        \subfloat[$\alpha$-plot requests]{\includegraphics[width=0.5\textwidth]{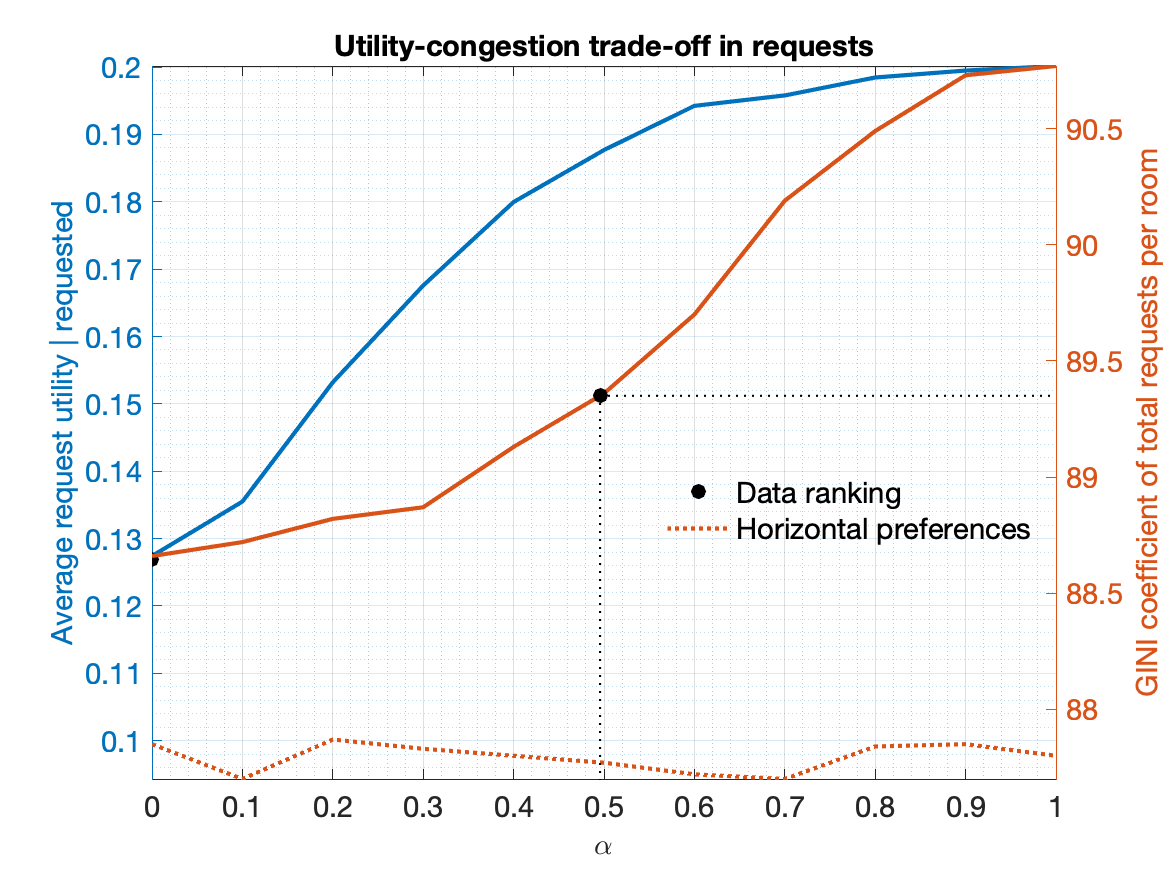}}
        \subfloat[Frontier requests]{\includegraphics[width=0.5\textwidth]{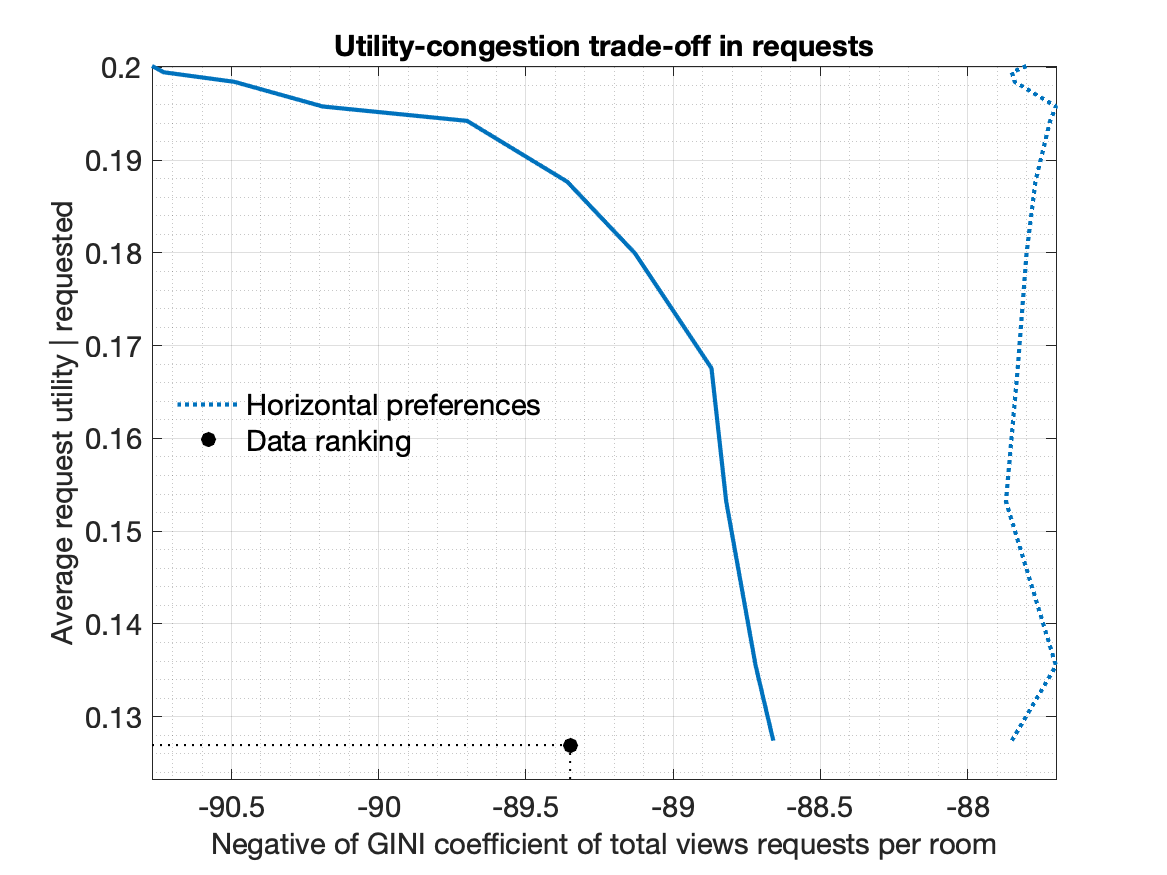}}
        \caption{Utility and congestion trade-off of clicks and requests with clusters\label{fig:9}}
    \end{figure}

\autoref{fig:9} replicates the same graphs in \autoref{fig:7} and \autoref{fig:6} in the main text for the case in which preferences are allowed to vary by clusters. Altogether, the qualitative results we find in the main text remain: (i) both the average utility of requested rooms and congestion (in clicks and requests) increase as we incorporate information of user preferences to a random ranking (i.e., as $\alpha$ moves from 0 to 1); (ii) the levels of utility and congestion obtained from the platform's algorithm are inefficient, in that we can increase utility (or reduce congestion) without increasing congestion (or reducing utility); (iii) if we ``garble'' the IDs of rooms in the data to destroy the correlation of preferences across users, the trade-off vanishes: increasing personalzation leads to a higher utility without increasing congestion. 

The key insight from \autoref{fig:9} is that the possibility frontier between congestion and utility becomes ``more concave'' as we allow preferences to vary across clusters. This is due to the fact that the model with clusters allows for more highly horizontally differentiated preferences since users in different clusters have different preferences for the same room (even if they happen to have the same individual characteristics). Therefore, introducing personalization to an otherwise random ranking (when $\alpha$ is close to zero) increases utility ``quicker'' (there is better targeting since the model approximates better the horizontal component of preferences) without increasing congestion ``too much'' (even if we personalize, users in different clusters still have different preferences for the same rooms). Altogether, this shows how the degree of horizontal and vertical differentiation in preferences mediates the trade-off between utility and congestion.
    
    \begin{table}[htbp!]
    \centering\small
    \caption{Parameter estimates of request utility by clusters\label{tab:7}}
    \begin{tabular}{lccc}\hline\hline
                                             &   Cluster 1   &   Cluster 2   &   Cluster 3   \\\hline
    Price                                    &     -0.0012   &     -0.0010** &     -0.0008   \\
                                             &    (0.0007)   &    (0.0005)   &    (0.0005)   \\
    Missing number of tenants                &     -0.0161   &     -0.1766   &     -0.0186   \\
                                             &    (0.1778)   &    (0.1328)   &    (0.1269)   \\
    Number of tenants                        &      0.0131   &     -0.0044   &     -0.0091   \\
                                             &    (0.0347)   &    (0.0250)   &    (0.0276)   \\
    Days since first published               &     -0.0005*  &     -0.0003*  &     -0.0001   \\
                                             &    (0.0003)   &    (0.0002)   &    (0.0002)   \\
    User has gender preferred by landlord    &      0.4040***&      0.4131** &      0.5149***\\
                                             &    (0.1248)   &    (0.1698)   &    (0.0710)   \\
    User has age preferred by landlord       &     -0.0060   &     -0.0377   &      0.0522   \\
                                             &    (0.1172)   &    (0.0907)   &    (0.0826)   \\
    User has occupation preferred by landlord&      0.0676   &     -0.0450   &      0.2497   \\
                                             &    (0.1106)   &    (0.1955)   &    (0.2071)   \\
    Has doorman                              &     -0.1683   &     -0.1185   &     -0.0664   \\
                                             &    (0.1209)   &    (0.0906)   &    (0.0873)   \\
    Has dishwasher                           &      0.0217   &      0.0458   &      0.1263*  \\
                                             &    (0.1124)   &    (0.0794)   &    (0.0767)   \\
    Has terrace                              &      0.1532   &     -0.0495   &     -0.0481   \\
                                             &    (0.1044)   &    (0.0712)   &    (0.0705)   \\
    Has AC                                   &      0.0871   &     -0.0056   &     -0.0057   \\
                                             &    (0.1064)   &    (0.0751)   &    (0.0753)   \\
    Is smoker friendly                       &      0.0115   &     -0.0620   &      0.0342   \\
                                             &    (0.0915)   &    (0.0661)   &    (0.0622)   \\
    Has elevator                             &      0.2132** &     -0.0245   &      0.0170   \\
                                             &    (0.0921)   &    (0.0679)   &    (0.0660)   \\
    Has exterior view                        &     -0.1442   &      0.0349   &     -0.0028   \\
                                             &    (0.0878)   &    (0.0657)   &    (0.0628)   \\
    Has TV                                   &     -0.1502*  &     -0.0484   &      0.0915   \\
                                             &    (0.0839)   &    (0.0607)   &    (0.0586)   \\
    Has balcony                              &      0.0605   &      0.0150   &      0.2056***\\
                                             &    (0.0843)   &    (0.0626)   &    (0.0596)   \\
    Has heating                              &      0.1849** &      0.0793   &      0.0666   \\
                                             &    (0.0912)   &    (0.0640)   &    (0.0619)   \\
    In Saint-Gervasi or Gracia               &     -0.2117   &     -0.0729   &     -0.0296   \\
                                             &    (0.3994)   &    (0.3069)   &    (0.3370)   \\
    In Ciutat Vella or Saint Marti           &      0.1377   &     -0.0257   &     -0.3318   \\
                                             &    (0.3882)   &    (0.2978)   &    (0.3343)   \\
    In Eixample                              &     -0.0425   &      0.0056   &     -0.1619   \\
                                             &    (0.3835)   &    (0.2971)   &    (0.3320)   \\
    In North of Barcelona                    &     -0.6274   &     -0.4368   &     -0.6578** \\
                                             &    (0.4110)   &    (0.3009)   &    (0.3349)   \\
    In Sants-Montjuic of Les Corts           &     -0.0807   &     -0.0631   &     -0.3314   \\
                                             &    (0.3771)   &    (0.2937)   &    (0.3231)   \\\hline
    Num Searches                             &      10,988   &      20,979   &      20,770   \\
    Num Results                              &      18,481   &      35,839   &      35,304   \\
    Pseudo-R2                                &      0.0271   &      0.0079   &      0.0260   \\
    Log-likelihood                           &   -990.4105   &  -1963.7888   &  -2062.3082   \\\hline\hline
    \multicolumn{4}{p{12cm}}{\scriptsize\textit{Notes:} Standard errors in parentheses (do not account for clustering stage); *$p<0.10$, **$p<0.05$, ***$p<0.01$. The category ``In Greater Barcelona'' is omitted.}
    \end{tabular}
    \end{table}

    \begin{table}[htbp!]
    \centering\small
    \caption{Parameter estimates of click propensity by clusters\label{tab:8}}
    \begin{tabular}{lccc}\hline\hline
                                          &   Cluster 1   &   Cluster 2   &   Cluster 3 \\\hline
    Position                              &     -0.0522***&     -0.0394***&     -0.1009***\\
                                          &    (0.0107)   &    (0.0081)   &    (0.0081)   \\
    Position$^2$                          &      0.0004   &      0.0008** &      0.0017***\\
                                          &    (0.0005)   &    (0.0003)   &    (0.0004)   \\
    1(Position = 1)                       &      0.1072** &      0.1421***&      0.0487   \\
                                          &    (0.0466)   &    (0.0335)   &    (0.0338)   \\
    1(Position = 2)                       &      0.0522   &      0.0009   &     -0.0326   \\
                                          &    (0.0399)   &    (0.0289)   &    (0.0288)   \\
    1(Position = 3)                       &      0.0237   &     -0.0027   &     -0.0168   \\
                                          &    (0.0348)   &    (0.0252)   &    (0.0250)   \\
    $\mathbb{E}[u|I,cl]$                  &      0.4379***&      0.1230   &      0.2219***\\
                                          &    (0.0684)   &    (0.0756)   &    (0.0621)   \\
    $\mathbb{E}[u|I,cl]$ $\times$ Position&      0.0069   &      0.0531***&      0.0262***\\
                                          &    (0.0066)   &    (0.0078)   &    (0.0059)   \\\hline
    Num Searches                          &      36,440   &      69,159   &      67,624   \\
    Num Results                           &     443,428   &     810,435   &     812,284   \\
    Pseudo-R2                             &      0.0156   &      0.0116   &      0.0220   \\
    Log-likelihood                        &  -3.498e+04   &  -6.487e+04   &  -6.422e+04   \\
    \hline\hline
    \multicolumn{4}{p{10cm}}{\scriptsize\textit{Notes:} Standard errors in parentheses (do not account for clustering or request stage); *$p<0.10$, **$p<0.05$, ***$p<0.01$.}
    \end{tabular}
    \end{table}

\clearpage

\subsection{Additional tables}\label{appendix:more_tables}

\autoref{tab:norm_req_bs} and \autoref{tab:norm_click_bs} report the normalized coefficients of the request utility and click propensity in the main text. More specifically, the parameters in \autoref{tab:norm_req_bs} correspond to the parameters in \autoref{tab:req_bs}, and the parameters in \autoref{tab:norm_click_bs} correspond to those in \autoref{tab:click_bs}.

    \begin{table}[htbp!]
    \vspace{1cm}
    \centering\small
    \caption{Normalized parameter estimates of request utility\label{tab:norm_req_bs}}
    \begin{tabular}{lcccc}\hline\hline
                                             &         (1)   &         (2)   &         (3)   &         (4)   \\\hline
    Price                                    &       -1.00&       -1.00&       -1.00&       -1.00\\
    Missing number of tenants                &            &     -130.93&      -96.72&      -85.94\\
    Number of tenants                        &            &        0.53&        1.13&       -4.02\\
    Days since first published               &            &       -0.40&       -0.32&       -0.26\\
    User has gender preferred by landlord    &            &      708.38&      603.54&      520.57\\
    User has age preferred by landlord       &            &       -9.85&        3.38&       10.28\\
    User has occupation preferred by landlord&            &      112.34&       94.65&       86.36\\
    Has doorman                              &            &            &     -113.70&     -120.95\\
    Has dishwasher                           &            &            &       89.54&       78.72\\
    Has terrace                              &            &            &      -23.48&      -14.73\\
    Has AC                                   &            &            &       10.03&       16.12\\
    Is smoker friendly                       &            &            &        2.10&       -2.08\\
    Has elevator                             &            &            &       65.10&       44.97\\
    Has exterior view                        &            &            &      -31.93&      -17.42\\
    Has TV                                   &            &            &      -14.12&      -13.54\\
    Has balcony                              &            &            &      141.31&      113.34\\
    Has heating                              &            &            &      109.70&      102.14\\
    In Saint-Gervasi or Gracia               &            &            &            &      -76.72\\
    In Ciutat Vella or Saint Marti           &            &            &            &     -109.53\\
    In Eixample                              &            &            &            &      -61.97\\
    In North of Barcelona                    &            &            &            &     -600.84\\
    In Sants-Montjuic of Les Corts           &            &            &            &     -183.20\\
    \hline\hline
    \multicolumn{5}{p{12cm}}{\scriptsize\textit{Notes:} Coefficients normalized to be in euros (divided by absolute value of price coefficient). The category ``In Greater Barcelona'' is omitted.}
    \end{tabular}
    \end{table}

    \begin{table}[hbtp!]
    \centering\small
    \vspace{1cm}
    \caption{Normalized parameter estimates of click propensity\label{tab:norm_click_bs}}
    \begin{tabular}{lcc}\hline\hline
                             &         (1)   \\\hline
    Position                 &     -249.14\\
    Position$^2$             &        3.74\\
    1(Position = 1)          &      352.42\\
    1(Position = 2)          &       -3.70\\
    1(Position = 3)          &       -9.68\\
    $\mathbb{E}[u|I]$        &        1.00\\
    $\mathbb{E}[u|I]$ $\times$ Position&      114.38\\\hline\hline
    \multicolumn{2}{p{5cm}}{\scriptsize\textit{Notes:} Coefficients normalized to be in euros (divided by absolute value of the normalized utility coefficient).}
    \end{tabular}
    \end{table}

\end{document}